\documentclass[journal]{IEEEtran}
\usepackage{cite}
\usepackage{graphicx}
\usepackage{amsmath}
\usepackage{times}
\usepackage{latexsym}
\usepackage{graphicx}
\usepackage{bm}
\usepackage{amssymb}
\usepackage[center]{caption2}
\usepackage{stfloats}
\usepackage{cases}
\usepackage{url}
\usepackage{array}
\usepackage{setspace}
\usepackage{fancyhdr}
\usepackage{braket}
\usepackage{color}
\usepackage{subfigure}
\usepackage{algorithm}
\usepackage{algorithmicx}
\usepackage{algpseudocode}
\usepackage{amsmath}
\DeclareMathAlphabet{\mathscr}{OT1}{pzc}{m}{it}

\newtheorem{theorem}{Theorem}

\newtheorem{corollary}{Corollary}

\newtheorem{remark}{Remark}
\def\proof{\noindent\hspace{2em}{\itshape Proof: }}
\def\endproof{\hspace*{\fill}~$\square$\par\endtrivlist\unskip}

\allowdisplaybreaks[4]
\newcommand{\RNum}[1]{\uppercase\expandafter{\romannumeral #1\relax}}

\begin{document}
\title{Integrated Sensing and Communication in IRS-assisted High-Mobility Systems: Design, Analysis and Optimization}
\author{Xingyu Peng, Qin Tao, Xiaoling Hu, Richeng Jin, Chongwen Huang, and Xiaoming Chen
\thanks{X. Peng, R. Jin, C. Huang, and X. Chen are with College of Information Science and Electronic Engineering, Zhejiang University, Hangzhou 310027, China, and Zhejiang Provincial Key Laboratory of Info. Proc., Commun. \& Netw. (IPCAN), Hangzhou 310027, China.  (Email: $\{$peng$\_$xingyu, richengjin, chongwenhuang, chen$\_$xiaoming$\}$@zju.edu.cn)
}
\thanks{Qin Tao is with the School of Information Science and Technology, Hangzhou Normal University, Hangzhou 311121, China. (Email: taoqin@hznu.edu.cn).}
\thanks{Xiaoling Hu is with the State Key Laboratory of Networking and Switching Technology, Beijing University of Posts and Telecommunications, Beijing 100876, China (email: xiaolinghu@bupt.edu.cn).}
}

\maketitle

\begin{abstract}
In this paper,  we investigate integrated sensing and communication (ISAC) in high-mobility systems with the aid of an intelligent reflecting surface (IRS). To exploit the benefits of Delay-Doppler (DD) spread caused by high mobility, orthogonal time frequency space (OTFS)-based frame structure and transmission framework are proposed. {In such a framework,} we first design a low-complexity ratio-based sensing algorithm for estimating the velocity of mobile user. Then, we analyze the performance of sensing and communication in terms of achievable mean square error (MSE)  and achievable rate, respectively, and reveal the impact of key parameters. Next, with the derived performance expressions, we jointly optimize the phase shift matrix of IRS and the receive combining vector at the base station (BS) to improve the overall performance of integrated sensing and communication. Finally, extensive simulation results confirm the effectiveness of the proposed algorithms in high-mobility systems.
\end{abstract}
\section{Introduction}
\subsection{Background}
Integrated sensing and communication (ISAC) has been identified as one of usage scenarios of sixth-generation (6G) wireless networks \cite{ISAC1,ISAC2,ISAC3}. The ascent of ISAC is propelled by its capacity to meet the demands of advanced applications like the metaverse, intelligent robotics, and virtual reality, that require both high-speed data transmission and high-precision sensing \cite{ISAC1}. Recently, ISAC also exhibits huge potential in Internet of Vehicles (IoV).  On the one hand, IoV needs to sense various information for vehicle coordination. On the other hand, IoV requires the support of communication for vehicle networking. Hence, ISAC presents an effective solution to IoV by integrating sensing and communication under the same hardware and spectrum resources \cite{ISAC4}. However, due to high-mobility of vehicles, there exists severe Delay-Doppler (DD) spread, degrading the performance of ISAC. It is imperative to design enabling ISAC techniques for high-mobility systems.

Orthogonal time frequency space (OTFS) modulation is known for its adaptability in high-mobility scenarios and spectral efficiency. In contrast to orthogonal frequency division multiplexing (OFDM), which operates in the time-frequency (TF) domain, OTFS modulates information symbols in the DD domain. To this end,  OTFS modulation, compared with OFDM, offers robustness against delay and Doppler shifts, fully exploiting channel diversity, and ensuring reliable communication \cite{embedded1}. Most importantly, the intrinsic ability of OTFS is to interact directly between transmitted signals and DD domain channel characteristics, i.e., delay and Doppler shifts, aligning seamlessly with sensing parameters such as distance and velocity \cite{potential}. Therefore, OTFS modulation is emerging as a promising candidate for ISAC in high-mobility systems.

Currently, ISAC is shifting towards higher frequency bands, such as millimeter-wave (mmWave) and terahertz (THz) for high-precision sensing and wideband communication. Yet, signal operating in mmWave or THz band faces challenges like significant molecular absorption, making line-of-sight (LoS) paths vulnerable to obstructions \cite{mmWave}. Fortunately, intelligent reflecting surface (IRS) provides an innovative solution. Firstly, IRS can be flexibly deployed on scatterers, thus creating a virtual LoS (VLoS) path between the transceivers. Moreover, IRS, comprising numerous low-cost reflecting elements, can not only reconfigure the transmission environment but also provide substantial passive gain \cite{IRS}. Importantly, IRS can mitigate performance degradation in OTFS systems arising from low Doppler resolution. The limited Doppler resolution leads to energy dispersion in the Doppler domain, negatively impacting communication and sensing efficacy. Utilizing time-varying, non-frequency-selective IRSs can generate virtual Doppler shifts, effectively countering the detrimental effects of fractional Doppler shifts due to finite resolution. This adjustment renders the effective Doppler shifts of cascaded channels distinct in the Doppler domain, thereby enhancing the performance of OTFS in high-mobility scenarios \cite{IRS-OTFS1}.
Furthermore, the phase rotation of the IRS, without causing additional delays, enhances path diversity in the Delay-Doppler (DD) domain, offering further improvements in system performance \cite{strongest}.


\subsection{Related Work}
Existing research on the topic of IRS-assisted ISAC systems includes two types of tasks: the design of the beamforming schemes \cite{IRS-ISAC1,IRS-ISAC2,IRS-ISAC3,IRS-ISAC4,IRS-ISAC5,IRS-ISAC6,IRS-IASC-vehicular1,IRS-IASC-vehicular2} and the estimation of sensing-related parameters \cite{TLS-ESPRIT,mmWave,estimation3,estimation4,estimation5}.  IRS can be deployed to enhance ISAC system performance through the design of joint active and passive beamforming schemes. For instance, to exploit the potential of IRS in supporting ISAC systems, the work\cite{IRS-ISAC1} proposed a joint beamforming optimization scheme to simultaneously detect a target and communicate with a user. This concept was expanded in \cite{IRS-ISAC2} to multi-user and multi-target scenarios by employing a penalty-based algorithm for beamforming design at the radar-communication (Radcom) base station (BS). The work\cite{IRS-ISAC6} then focused on the IRS-assisted ISAC system in cluttered environments and proposed a low-complexity beamforming scheme to minimize the interference between sensing and communication signals. The work \cite{IRS-IASC-vehicular1} explored the application of a simultaneous transmitting and reflecting surface in vehicular networks, where an end-to-end multi-agent deep reinforcement learning framework was introduced to address the challenges of optimizing joint sensing and communication. The work \cite{IRS-IASC-vehicular2} investigated the application of intelligent omni-surfacing in vehicular networks, addressing the joint beamforming optimization problem under the impact of location estimation errors.
Furthermore, the work \cite{IRS-ISAC5} investigated the integration of multiple IRSs into ISAC systems, revealing their potential for significant beamforming gains and enhanced system performance compared to a single IRS. 
Besides, IRS can also provide additional spatial degrees of freedom and enhance the accuracy of the estimated sensing-related parameters. For example, the work \cite{TLS-ESPRIT} and \cite{mmWave} demonstrated centimeter-level localization accuracy in ISAC systems using two semi-passive IRSs, validating the effectiveness of IRS in assisting the estimation of the user's location as well as the communication. 
The work \cite{estimation3} explored joint communication and velocity estimation by estimating the Doppler shifts of the sensing channel. Additionally, the work \cite{estimation4} predicted the areas of interest (AoI) based on user trajectory analysis, including direction, distance, and rate of directional change. However, these works predominantly concentrate on low-mobility scenarios, potentially limiting their applicability to high-mobility scenarios.

As an emerging modulation technique, OTFS has gained prominence for its robustness against delay and Doppler shifts, especially in high-mobility scenarios. Its superior performance in demodulation, as shown in \cite{basic}, establishes OTFS as a more effective modulation scheme than OFDM. Additionally, OTFS boasts a lower peak to average power ratio (PAPR), enhanced spectral efficiency, and better adaptability in high-mobility scenarios compared to OFDM \cite{PAPR}. Therefore, OTFS has been advocated as a promising scheme for ISAC systems, especially in high-mobility scenarios \cite{OTFS-ISAC1,OTFS-ISAC2,OTFS-ISAC3,OTFS-ISAC4,OTFS-ISAC7}.
Notably, the work \cite{OTFS-ISAC3} explored a joint radar parameter estimation and communication system using OTFS modulation, assessing the effectiveness of the OTFS-ISAC system. This approach was further advanced in \cite{OTFS-ISAC7}, where multiple-input multiple-output (MIMO) systems were considered, and a maximum likelihood (ML) detector and hybrid beamforming were designed for reliable communication and accurate estimation.
Moreover, the work \cite{OTFS-ISAC2} designed a spatially spread OTFS-based ISAC framework, realizing beam tracking for radar detection and precoding optimization at the BS. This work uniquely derived the pair-wise error probability (PEP) for communication, which is leveraged for both precoding and power allocation design.
The work \cite{OTFS-ISAC1} proposed a novel delay-Doppler-angle estimation method in MIMO-OTFS systems by employing a generalized likelihood ratio test, which effectively mitigates and exploits inter-symbol interference (ISI) and inter-carrier interference (ICI), enhancing target detection capabilities in range and velocity. However, it primarily focused on the trade-off between achievable rate and radar signal-to-noise ratio (SNR), rather than the estimation accuracy of the proposed method.  

IRS can seamlessly integrate with OTFS systems, offering significant flexibility in system design. Through the deployment of the IRS, system performance can be improved \cite{IRS-OTFS5}.
Current research on IRS-assisted OTFS systems primarily focuses on two aspects: IRS phase shift design for performance enhancement and channel estimation \cite{embedded1,strongest,IRS-OTFS2,IRS-OTFS3,IRS-OTFS4,IRS-OTFS7,IRS-OTFS8,SBL_MUSIC,IRS-IASC-vehicular3,IRS-IASC-vehicular4,tao2}. 
In particular, the work \cite{embedded1} proposed an OTFS frame structure with embedded pilot symbols so that the delay and Doppler can be estimated directly in the DD domain. Then, the work \cite{SBL_MUSIC} extended to the scenarios with Doppler spread caused by fractional Doppler shifts.
The work \cite{IRS-IASC-vehicular4} estimated channel 
state information (CSI) between the vehicles and BS using the multidimensional orthogonal matching pursuit (MOMP) approach. 
Moreover, the work \cite{IRS-OTFS2} and \cite{IRS-OTFS3} proposed a method for joint channel estimation and data detection, based on the assumption that each delay tap only correlates to a singular Doppler shift.
The work \cite{IRS-IASC-vehicular3} explored the ISAC in the vehicle network via STARS, by utilizing both the angular and Doppler information of the channel, the IRS phase shift matrix was designed for vehicular tracking.
To simplify the optimization problem, \cite{strongest, IRS-OTFS7, IRS-OTFS8,IRS-IASC-vehicular3,IRS-IASC-vehicular4} proposed a beamforming design that aligns only with the strongest delay-Doppler path, typically the LoS path among the BS, the IRS, and the user, while ignoring the impact of other paths.

However, practical exploration of IRS-assisted OTFS modulation systems remains limited. Current beamforming schemes in IRS-OTFS works tend to be overly simplistic, either by aligning IRS phase shifts solely with the strongest path, overlooking interference from other paths, or by assuming a one-to-one correspondence between delay taps and Doppler shifts. Furthermore, in high-mobility environments, precise velocity estimation becomes critical for tasks such as vehicle positioning prediction and beam tracking, necessitating high-accuracy and low complexity algorithms that can adapt to rapid channel variations.
Therefore, there is a crucial need to delve into IRS-assisted ISAC systems with high mobility, involving low-complexity methods that estimate the sensing-related parameters and beamforming design schemes based on theoretical analyses of sensing methods to fully exploit the potential of IRS.

\subsection{Contributions}
In this paper, we focus on an IRS-assisted ISAC system with high mobility. We provide a low-complexity method for estimating the sensing-related parameters and develop joint beamforming at the BS and IRS that leverages the path diversity to effectively balance communication and sensing performance.
The main contributions are summarized as follows:
\begin{itemize}
\item{We propose an OTFS-based framework for ISAC in IRS-assisted high-mobility systems under the same hardware and spectrum resources.}
\item{We develop a low-complexity ratio-based sensing algorithm to estimate the velocity of mobile user and derive its performance in terms of effective sensing probability and achievable mean square error. }
\item{We design a subspace-based joint IRS phase shift matrix and BS receive combining vector optimization algorithm to improve the overall performance of sensing and communication in high-mobility scenarios.}
\end{itemize}

The remainder of the paper is organized as follows. Section \ref{se:se2} introduces the IRS-assisted ISAC system and the OTFS frame. The ratio-based sensing algorithm and its performance are presented in Section \ref{se:se3}, while the subspace-based beamforming design algorithm is introduced in Section \ref{se:se4}. Numerical results and discussions are provided in Section \ref{se:se5}, and finally, Section \ref{se:se6} concludes the paper.

\section{System Model}\label{se:se2}
\begin{figure}[htbp]
\centering
\includegraphics[width=3.5in]{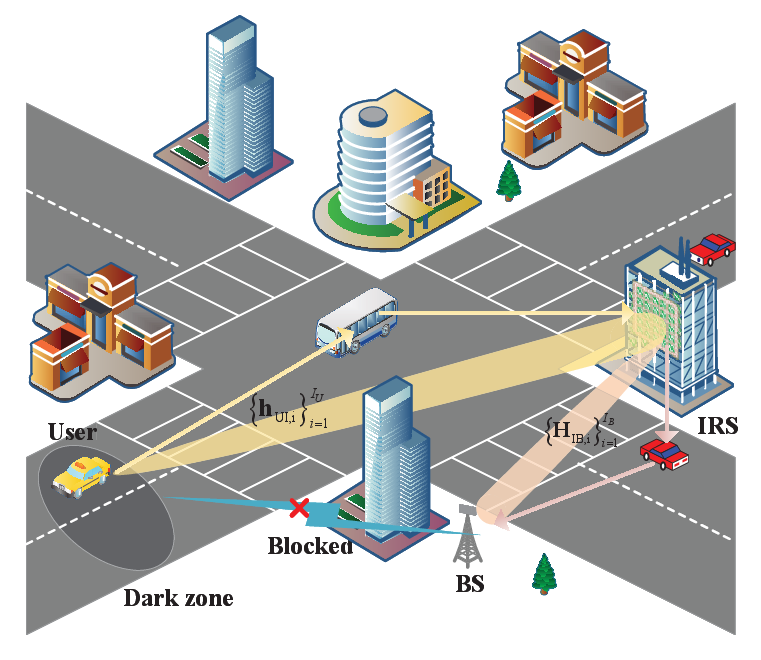}
\caption{System model: An IRS-assisted high-mobility system.}\label{fig:fig1}
\end{figure}

Consider a mobile communication system, as illustrated in Fig. \ref{fig:fig1}, where a single-antenna mobile user\footnote{The proposed scheme can be extended to multiuser scenarios directly by allocating orthogonal resource blocks to the users \cite{embedded1}. Furthermore, it is applicable in scenarios where the IRS is user-mounted, directly.} communicates with a BS with $N_B$-element uniform linear array (ULA) under the mmWave spectrum band. The paths between the BS and the user are obstructed due to the weak penetrability of mmWave, and an IRS equipped with $N_I=N_{I,1} \times N_{I,2}$ reflecting elements is deployed to establish a solid VLoS path to assist the communication \footnote{Assuming the user-IRS association has been established through the methodologies described in \cite{asso1,asso2,asso3}.}. Based on the received signal sent from the mobile user, the BS senses the velocity of the mobile user and recovers the data information simultaneously.  
\subsection{Time Domain Channel Model}
In the case of high mobility, the time domain channel from the user to IRS can be expressed as \cite{IRS-OTFS2}\vspace{0mm}
\begin{equation}\vspace{0mm}
\begin{aligned}
{\bf{h}}_{{\mathrm{UI},i}}(t)
&= \sum\limits_{p = 1}^{{L_{{\rm{UI}}}}} {h_p^{{\rm{UI}}}} {e^{\jmath 2\pi v_p^{{\rm{UI}}}\left( {t - \tau _p^{{\rm{UI}}}} \right)}}\delta \left( {i{T_s} - \tau _p^{{\rm{UI}}}} \right){{\bf{a}}_{\rm{I}}}\left( {\phi _p^{{\rm{UI}}},\psi _p^{{\rm{UI}}}} \right),\\ &\qquad\qquad\qquad\qquad\qquad i = 0, \ldots ,{{\rm{I}}_{\rm{U}}} - 1,
\end{aligned}
\end{equation}
where ${L_{{\rm{UI}}}}$ is the number of paths between the user and the IRS, ${h_p^{{\rm{UI}}}}$, $v_p^{{\rm{UI}}}$, and $\tau _p^{{\rm{UI}}}$ are complex channel gain, Doppler shifts, delay of $p$-th path, respectively. $\delta(\cdot)$ is the Delta function, $i$ denotes the delay domain index, ${{\rm{I}}_{\rm{U}}}$ is the maximum number of delay taps, and $T_s$ is the system sampling period.  Moreover,  $\phi _p^{{\rm{UI}}}$ and $ \psi _p^{{\rm{UI}}}$ are the elevation and azimuth effective angles of arrival (AoAs) at the IRS, ${{\bf{a}}_{\rm{I}}}\left( {\phi _p^{{\rm{UI}}},\psi _p^{{\rm{UI}}}} \right)$ is the array response vector at the IRS, which is given by\vspace{0mm}
\begin{equation}\vspace{0mm}
\begin{aligned}
\mathbf{a}_{\rm{I}}(u,v) & ={{\left[ 1,\cdots ,{{e}^{j(i_1-1)u}},\cdots ,{{e}^{j\left( {{N}_{I,1}}-1 \right)u}} \right]}^{T}}\\
&\qquad \quad\otimes {{\left[ 1,\cdots ,{{e}^{j(i_2-1)v}},\cdots ,{{e}^{j\left( {{N}_{I,2}}-1 \right)v}} \right]}^{T}}, \\
&\qquad\qquad i_1=1,\cdots,{{N}_{I,1}},i_2=1,\cdots,{{N}_{I,2}},
\end{aligned}
\end{equation}
where $j$ denotes the imaginary part, and  $\otimes$ denotes the Kronecker product.

Similarly, the time domain channel from the IRS to BS can be expressed as\vspace{0mm}
\begin{equation}\vspace{0mm}
\begin{aligned}
{{\bf{H}}_{{\mathrm{IB}},i}}(t) & = \sum\limits_{p = 1}^{{L_{{\rm{IB}}}}} {h_p^{{\rm{IB}}}} {e^{\jmath 2\pi v_p^{{\rm{IB}}}\left( {t - \tau _p^{{\rm{IB}}}} \right)}}\delta \left( {i{T_s} - \tau _p^{{\rm{IB}}}} \right){{\bf{a}}_{\rm{B}}}\left( {\vartheta _p^{{\rm{IB}}}} \right)\\
&\qquad\qquad{{\bf{a}}_{\rm{I}}}\left( {\phi _p^{{\rm{IB}}},\psi _p^{{\rm{IB}}}} \right),i = 0, \ldots ,{{\rm{I}}_{\rm{B}}} - 1,
\end{aligned}
\end{equation}
where ${L_{{\rm{IB}}}}$ is the number of paths between the IRS and the BS.  ${{\rm{I}}_{\rm{B}}}$ is the maximum number of delay taps, ${h_p^{{\rm{IB}}}}$, $v_p^{{\rm{IB}}}$,  $\tau _p^{{\rm{IB}}}$,  ${\vartheta _p^{{\rm{IB}}}}$, $\phi _p^{{\rm{IB}}} $ and $\psi _p^{{\rm{IB}}}$ are complex channel gain, Doppler shifts, delay, the effective AoA at the BS, elevation and azimuth effective angles of departure (AoDs) at the IRS of $p$-th path, respectively.
${{\bf{a}}_{\rm{B}}}\left( {\vartheta _p^{{\rm{IB}}}} \right)$ is the array response vector at the BS, which is given by\vspace{0mm}
\begin{equation}\vspace{0mm}
\begin{aligned}
 {{\bf{a}}_{\rm{B}}}\left( {\vartheta _p^{{\rm{IB}}}} \right) & ={{\left[ 1,\cdots ,{{e}^{j(i_3-1){\vartheta _p^{{\rm{IB}}}} }},\cdots ,{{e}^{j\left( {{N}_{B}}-1 \right){\vartheta _p^{{\rm{IB}}}} }} \right]}^{T}},\\
&\qquad\qquad\qquad\qquad i_3=1,\cdots,{{N}_{\rm{B}}}
\end{aligned}
\end{equation}

\subsection{OTFS-based Transmission Framework}
\begin{figure*}[htbp]
\centering
\includegraphics[width=4.5in]{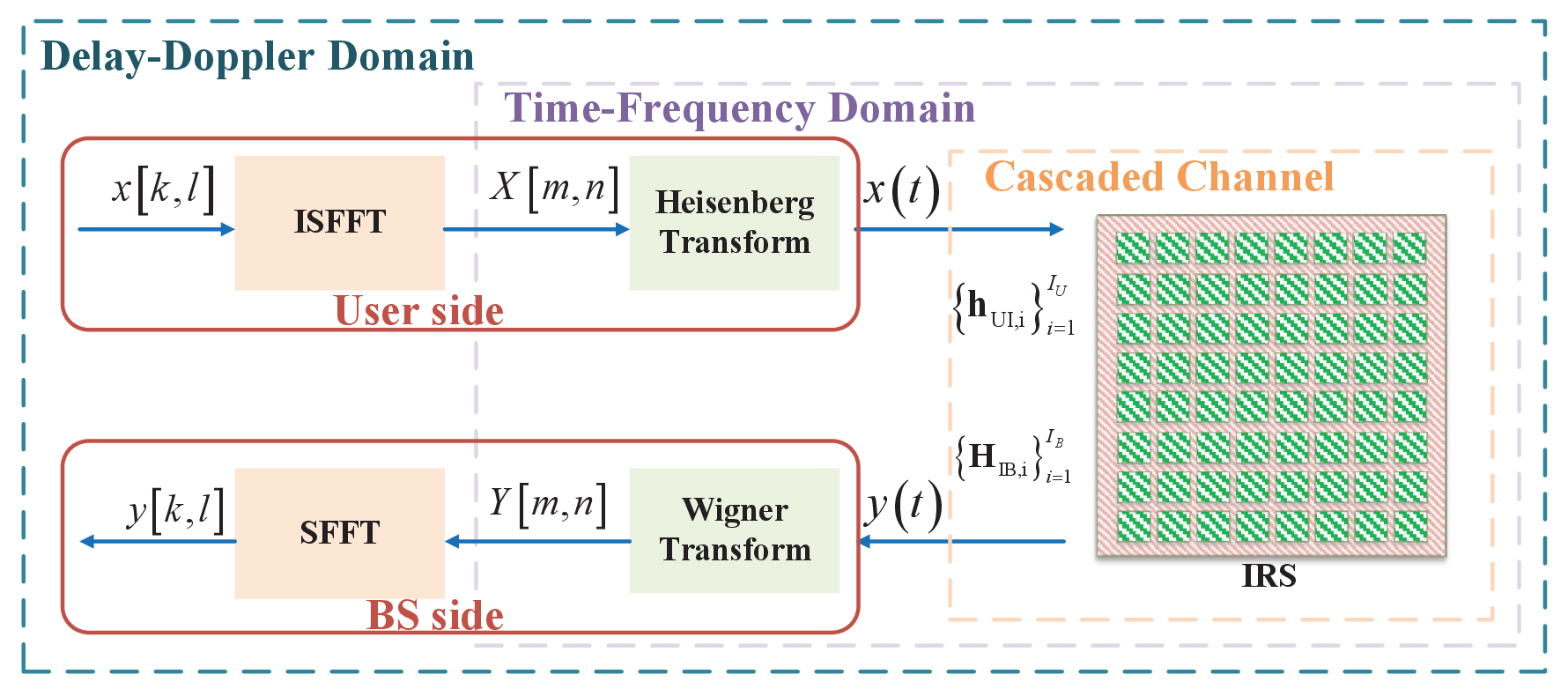}
\caption{{The OTFS-based transmission {framework}.}}\label{fig:fig2}
\end{figure*}
Considering severe Delay-Doppler spread of wireless channels in high-mobility scenarios, OTFS modulation is employed to improve the overall performance of sensing and communication. Specifically, in an OTFS frame, the information symbols ${\mathbf{X}^{\mathrm{DD}}[k,l]},\forall k=[0,\cdots,N-1],\forall l=[0,\cdots,M-1]$ are arranged in the DD domain, where the grid set is defined as $\mathcal{G}^{\mathrm{DD}}=\left\{\left.\left(\frac k{NT_s},\frac l{M\Delta f}\right)\right|k=0,\cdots, N-1,l=0,1,\cdots,M-1\right\}$, where $\Delta f=\frac{1}{T_s}$ is the subcarrier spacing. The duration of the OTFS frame is $T_f=NT_s$ and the sampling interval along the Doppler dimension is $\frac 1{NT_s}$. Moreover, $B=M \Delta f$ is the total bandwidth and $\frac{1}{M \Delta f}$  is the sampling interval along the delay dimension. With the transmission framework shown in Fig. \ref{fig:fig2}, the inverse symplectic finite Fourier transform (ISFFT) is first applied to convert ${\mathbf{X}^{\mathrm{DD}}[k,l]}$ into a block of samples in the TF domain ${\mathbf{X}^{\mathrm{TF}}[n,m]}$, which is given by\vspace{0mm}
\begin{equation}\vspace{0mm}
\begin{aligned}
&\mathbf{X}^{\mathrm{TF}}\left[ {n,m} \right] = \sum\limits_{k = 0}^{N - 1} {\sum\limits_{l = 0}^{M - 1} \mathbf{X}^{\mathrm{DD}} } \left[ {k,l} \right]{e^{j2\pi \left( {\frac{{nk}}{N} - \frac{{ml}}{M}} \right)}}, \\
&\qquad \qquad\forall n=[0,\cdots,N-1],\forall m=[0,\cdots,M-1].
\end{aligned}
\end{equation}
Then, Heisenberg transform is conducted to transform the TF domain samples to the following continuous-time signal\vspace{0mm}
\begin{equation}\vspace{0mm}
s\left( t \right) = \sum\limits_{n = 0}^{N - 1} {\sum\limits_{m = 0}^{M - 1} \mathbf{X}^{\mathrm{TF}} } \left[ {n,m} \right]{g_{{\rm{tx}}}}\left( {t - nT_s} \right){e^{j2\pi m\Delta f\left( {t - nT_s} \right)}},
\end{equation}
where the transmit shaping pulse ${g_{{\rm{tx}}}}$ is a rectangular function with duration $T_s$. The mobile user sends the signal $s\left( t \right)$ over the uplink channel, and thus, the IRS receives a noiseless signal as follows\vspace{0mm}
\begin{equation}\vspace{0mm}
{{\bf{y}}^{\rm{I}}}(t) 
= \sum\limits_{p = 1}^{{L_{{\rm{UI}}}}} {h_p^{{\rm{UI}}}} {e^{j2\pi v_p^{{\rm{UI}}}(t - \tau _p^{{\rm{UI}}})}}{{\bf{a}}_{\rm{I}}}\left( {\phi _p^{{\rm{UI}}},\psi _p^{{\rm{UI}}}} \right)s(t - \tau _p^{{\rm{UI}}}).
\end{equation}

Let $\boldsymbol{\Theta}=\mathrm{\mathrm{diag}}  \left(\boldsymbol{\xi}\right)$ \footnote{To simplify hardware implementation and reduce costs, we adopt the time-invariant IRS approach, where the phase shift remains constant within each OTFS frame but varies across different OTFS frames, ensuring that the IRS does not change the Doppler shifts of the cascade channel while enhancing overall system performance 
\cite{IRS-OTFS1,real-time-IRS}.}denote the phase shift matrix of the IRS, where $\boldsymbol{\xi}=\left [ e^{\mathrm{j}\theta _{1}},\cdots , e^{\mathrm{j}\theta _{N_I}} \right ] ^T$ is the phase shift beam and $\theta _{i}\in [0,2\pi], \forall i=[1,\cdots,N_I]$ is the phase shift of the $i$-th element of the IRS. Reflected by the IRS, the combined received signal at the BS can be expressed as\vspace{-3mm}
\begin{equation}\vspace{0mm}
\begin{aligned}
&{\bf{y}^{\rm{B}}}\left( t \right) = \mathbf{r}^H\sum\limits_{i_B=0}^{I_B-1}{{\bf{H}}_{{\rm{IB,i}}}}(t)\boldsymbol{\Theta}\mathbf{y}^I(t-i_BT_s) \\
 &=  {{\bf{r}}^H}{\sum\limits_{p_1 = 1}^{{L_{{\rm{UI}}}}}\sum\limits_{p_2 = 1}^{{L_{{\rm{IB}}}}} {h_{p_2}^{{\rm{IB}}}} h_{p_1}^{{\rm{UI}}}{e^{\jmath 2\pi \left( {v_{p_2}^{{\rm{IB}}} + v_{p_1}^{{\rm{UI}}}} \right)\left( {t - \tau _{p_2}^{{\rm{IB}}} - \tau _{p_1}^{{\rm{UI}}}} \right)}}{e^{\jmath 2\pi v_{p_2}^{{\rm{IB}}}\tau _{p_1}^{{\rm{UI}}}}}} \\
&\qquad{{\bf{a}}_{\rm{B}}}\left( {\vartheta _{p_2}^{{\rm{IB}}}} \right){\bf{a}}_{\rm{I}}^H\left( {\phi _{p_2}^{{\rm{IB}}},\psi _{p_2}^{{\rm{IB}}}} \right)\boldsymbol{\Theta} {{\bf{a}}_{\rm{I}}}\left( {\phi _{p_1}^{{\rm{UI}}},\psi _{p_1}^{{\rm{UI}}}} \right)  s(t - {\tau ^{{\rm{IB}}}} - \tau _{p_1}^{{\rm{UI}}}) ,
\end{aligned}
\end{equation}
where $\left \| \mathbf{r} \right \| =1$ denotes the combining vector at the BS and the superscript ``$\rm{H}$'' represents the Hermitian transpose. By using the Wigner transform, the received samples can be obtained as\vspace{0mm}
\begin{equation}\label{eq:eq9}\vspace{0mm}
\mathbf{Y}^{\mathrm{TF}}\left[ {n,m} \right] = \mathbf{Y}\left( {t,f} \right){|_{t = nT_s,f = m\Delta f}}, 
\end{equation}
where $
\mathbf{Y}\left( {t,f} \right) = \int {{\bf{y}^{\rm{}}}\left( t \right)} g_{{\rm{rx}}}^*\left( {t' - t} \right){e^{ - j2\pi ft'}}dt'$ and the superscript ``$*$'' represents the complex conjugate. Further, by applying the symplectic finite Fourier transform (SFFT) to (\ref{eq:eq9}), the received samples in the DD domain can be expressed as \vspace{0mm}
\begin{equation}\vspace{0mm}
{\mathbf{Y}^{\mathrm{DD}}}\left[ {k,l} \right] = \frac{1}{{NM}}\sum\limits_{n = 0}^{N - 1} {\sum\limits_{m = 0}^{M - 1} \mathbf{Y}^{\mathrm{TF}} } \left[ {n,m} \right]{e^{ - j2\pi \left( {\frac{{nk}}{N} - \frac{{ml}}{M}} \right)}}.
\end{equation} 

Define $\tau _{p_1}^{{\rm{UI}}}=\frac{l_{{\tau _{{p_1}}}}}{M\Delta f}$, $\tau _{p_2}^{{\rm{IB}}}=\frac{l_{{\tau _{{p_2}}}}}{M\Delta f}$, ${\nu_{{p_1}}^{{\rm{UI}}}}=\frac{\left({{k_{{\nu _{{p_1}}}}}}+{{\chi  _{{\nu _{{p_1}}}}}}\right)}{NT_s}$, and ${\nu_{{p_2}}^{{\rm{IB}}}}=\frac{\left({{k_{{\nu _{{p_2}}}}}}+{{\chi  _{{\nu _{{p_2}}}}}}\right)}{NT_s}$, where ${l_{{\tau _{{p_1}}}}^{\mathrm{UI}}}$ and ${l_{{\tau _{{p_1}}}}^{\mathrm{UI}}}$ denote the delay tap of $ \tau _{p_1}^{{\rm{UI}}}$ and $\tau _{p_2}^{{\rm{IB}}}$. {$\{{k_{{\nu _{{p_1}}}}},{k_{{\nu _{{p_2}}}}}\}$ and  $\{{\chi  _{{\nu _{{p_1}}}}},{\chi  _{{\nu _{{p_2}}}}}\}$ are respectively the integer tap and its related fractional bias of the Doppler shift $\{{\nu_{{p_1}}^{{\rm{UI}}}},{\nu_{{p_2}}^{{\rm{IB}}}}\}$.} The received signal in the DD domain can be rewritten as 
\begin{equation}\label{eq:eq11}\vspace{0mm}
\begin{aligned}
&{\mathbf{Y}^{\mathrm{DD}}}\left[ {k,l} \right] \\
& = \sum\limits_{k' = 0}^{N - 1} {\sum\limits_{l' = 0}^{M - 1} {\mathbf{X}^{\mathrm{DD}}\left[ {k',l'} \right]} } \sum\limits_{{p_2} = 1}^{{L_{{\rm{IB}}}}} {\sum\limits_{{p_1} = 1}^{{L_{{\rm{UI}}}}} {h_{p_2}^{{\rm{IB}}}} h_{p_1}^{{\rm{UI}}}{e^{\jmath 2\pi v_{p_2}^{{\rm{IB}}}\tau _{p_1}^{{\rm{UI}}}}}} {{\bf{r}}^H} {{\bf{a}}_{\rm{B}}}\left( {\vartheta _{p_2}^{{\rm{IB}}}} \right)\\
&\quad{\bf{a}}_{\rm{I}}^H\left( {\phi _{p_2}^{{\rm{IB}}},\psi _{p_2}^{{\rm{IB}}}} \right)\mathrm{diag}\left( {{{\bf{a}}_{\rm{I}}}\left( {\phi _{p_1}^{{\rm{UI}}},\psi _{p_1}^{{\rm{UI}}}} \right)} \right){\boldsymbol{\xi \Psi }}_{k,{k^\prime }}^{{p_1},{p_2}}\left[ {l,{l^\prime }} \right]  + \mathbf{N}\left[ {k,l} \right],
\end{aligned}
\end{equation}
where $\mathbf{N}\left[ {k,l} \right]$ denotes the additive white Gaussian noise (AWGN) with zero mean and covariance $\sigma^2$, and the matrix ${\boldsymbol{\Psi }}_{k,{k^\prime }}^{p_1,p_2}[l,l']$ is defined as\vspace{0mm}
\begin{equation}\vspace{0mm}
\begin{array}{l}
{\boldsymbol{\Psi }}_{k,{k^\prime }}^{p_1,{p_2}}[l,l'] = \frac{1}{N}\sum\limits_{n = 0}^{N - 1} {e^{j2\pi \left( {k' - k + \nu _{{p_1},{p_2}}^{{\rm{UI}}B}NT_s} \right)\frac{n}{N}}}{e^{ - j2\pi \nu _{{p_1},{p_2}}^{{\rm{UI}}B}\frac{{l'}}{{M\Delta f}}}}\\
\qquad\times \left\{ \begin{array}{*{20}{l}}
1&{{\rm{if    }}\;\;{l_{{\tau _{{{p_1}}}}}^\mathrm{UI}} + {l_{{\tau _{{{p_2}}}}}^\mathrm{IB}} \le l < M}\\
{{e^{ - j2\pi \left( {\frac{{k'}}{N} + \nu _{{p_1},{p_2}}^{{\rm{UI}}B}T_s} \right)}}}&{{\rm{if    }}\;\;0 \le l < {l_{{\tau _{{{p_1}}}}}^\mathrm{UI}} + {l_{{\tau _{{{p_2}}}}}^\mathrm{IB}}}
\end{array} \right. ,
\end{array}
\end{equation}
where ${l_{{\tau _{p_1,p_2}}}}={l_{{\tau _{{p_1}}}}^{\mathrm{UI}}}+{l_{{\tau _{{p_2}}}}^{\mathrm{IB}}}$, $\nu _{p_1,{p_2}}^{{\rm{UI}}B}={\nu_{{p_1}}^{{\rm{UI}}}}+{\nu_{{p_2}}^{{\rm{IB}}}}$, and $\tau _{p_1,{p_2}}^{{\rm{UI}}B}=\tau _{p_1}^{{\rm{UI}}}+\tau _{p_2}^{{\rm{IB}}}$.
%

\subsection{Symbol Arrangement in OTFS Frame}
\begin{figure}[t]
\centering
\includegraphics[width=3.5in]{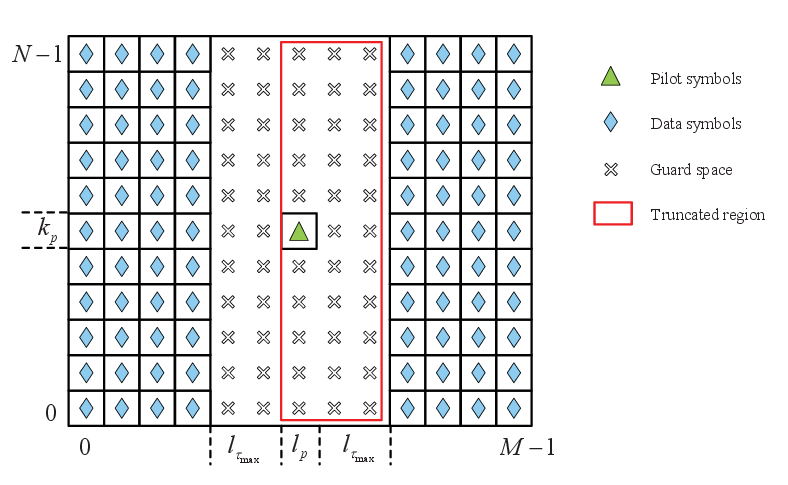}
\caption{The arrangement of pilot and data symbols in the OTFS frame.}\label{fig:fig3}
\end{figure}
In order to effectively integrate sensing and communication in high-mobility systems, pilot and data symbols are arranged in the OTFS according to a given pattern \cite{embedded1}, as shown in Fig. \ref{fig:fig3}. Specifically, a pilot symbol $x_p$ combined with transmitted power is inserted in the center of the DD domain, i.e., $x[k_p,l_p]=x_p$. Moreover, the guard space for mitigating interference from data symbols to pilot symbols is arranged in the range $k\in\left \{ {{\cal K}_{\mathrm{g}}}|0\leq k \leq N-1,k \neq k_p \right \} $ and $l\in \left \{ {{\cal L}_{\mathrm{g}}}|l_p-l_{\rm{max}}\le l \le l_p+l_{\rm{max}}, l \neq l_p \right \} $, where $l_{\mathrm{max}}=\lceil\tau_{\mathrm{max}}M\Delta f\rceil$\footnote{{It can be measured in advance, for example, in Long Term Evolution (LTE) channels, $l_{\mathrm{max}}$=20 as shown in \cite{LTE}.}},  $\tau_{\mathrm{max}}=\max\limits_{p_1,p_2}\{\tau _{p_1,{p_2}}^{{\rm{UI}}B}\}$, and $\lceil \cdot \rceil$ denotes the ceiling function which returns the smallest integer larger than the input value. Other than pilot and guard space, data symbols are inserted in the rest of the OTFS frame. Hence, the whole OTFS frame can be expressed as
\begin{equation}\label{eq:eq13}
\mathbf{X}^{\mathrm{DD}}[k,l]=\begin{cases}x_p,&k=k_p,l=l_p\\0,&l\in{{\cal L}_{\mathrm{g}}}, k\in {{\cal K}_{\mathrm{g}}}
\\x_d[k,l],&\text{otherwise}\end{cases}.
\end{equation}

Under this setup, the channel between the BS and the user can be estimated based on the symbols in the truncated region by using the existing method in \cite{embedded1} and \cite{embedded2}.

\section{Design of Velocity Sensing Algorithm}\label{se:se3}
In this section, we first design a sensing algorithm to estimate the velocity of mobile user\footnote{Since the locations of IRS and BS are relatively stable, the Doppler shift of the LoS path intuitively responds to the velocity of the user, i.e., $\nu_1^{{\rm{UI}}}=\frac{v_u}{\lambda}\cos \theta_u $, where $v_u$, $\lambda$, and $\theta_u$ denote the velocity of the user, the carrier wavelength, and the effective angle between the user and the IRS.}. Then, we analyze the performance of the proposed sensing algorithm in terms of effective sensing probability and achievable mean square error (MSE).

\subsection{Ratio-based Sensing Algorithm}
Due to Delay-Doppler spread, the inserted pilot symbol may be present from $l=l_p \cdots l_p+l_{\rm{max}}$ and $k=0 \cdots N-1$ in the received OTFS frame. Hence, the received pilot symbol can be expressed as
\begin{equation}
\begin{aligned}
&{\bf{Y}}^{\mathrm{DD}}_{\rm{p}}\left[ {k,l} \right] 
 = \frac{{{x_p}}}{N} \times\\
&\sum\limits_{n=0}^{N - 1} {\sum\limits_{{p_1},{p_2} = 1}^{{P_{{l_{{p_1},{p_2}}}}}} {{e^{\frac{{j2\pi {k_p}n}}{N}}}h_{{p_1},{p_2}}^{{\rm{UIB}}}{e^{j2\pi \left( {v_{{p_1}}^{{\rm{UI}}} + v_{{p_2}}^{{\rm{IB}}}} \right)T_sn}}{e^{\frac{{ - j2\pi kn}}{N}}}} } + \mathbf{N}\left[ {k,l} \right],
\end{aligned}
\end{equation}
where ${{P_{{l_{{p_1},{p_2}}}}}}$ is the number of multipath with delay ${l_{{\tau _{p_1,p_2}}}}=l-l_p$, 
\begin{equation}
{h_{{p_1},{p_2}}^{{\rm{UIB}}}}= {h_{{p_2}}^{{\rm{IB}}}h_{{p_1}}^{{\rm{UI}}}{e^{\jmath 2\pi v_{{p_2}}^{{\rm{IB}}}\tau _{{p_1}}^{{\rm{UI}}}}}} {{\beta _{{p_1},{p_2}}}{e^{\jmath 2\pi \left( {v_{{p_2}}^{{\rm{IB}}} + v_{{p_1}}^{{\rm{UI}}}} \right)\frac{{l - {l_{{\tau _{{p_1}}}}^{\mathrm{UI}}} - {l_{{\tau _{{p_2}}}}^{\mathrm{IB}}}}}{{M\Delta f}}}}},
\end{equation}
and
\begin{equation}
{{\beta _{{p_1},{p_2}}}} = {{{\bf{r}}^H}{{\bf{a}}_{\rm{B}}}\left( {\vartheta _{{p_2}}^{{\rm{IB}}}} \right){\bf{a}}_{\rm{I}}^H\left( {\phi _{{p_2}}^{{\rm{IB}}},\psi _{{p_2}}^{{\rm{IB}}}} \right)\boldsymbol{\Theta} {{\bf{a}}_{\rm{I}}}\left( {\phi _{p1}^{{\rm{UI}}},\psi _{p1}^{{\rm{UI}}}} \right)}.
\end{equation}

It is noticed that the LoS path, both from the user to the IRS and from the IRS to the BS, exhibits the minimum delay. This has been identified in existing research, such as \cite{embedded1}, leading to a reasonable assumption that ${{P_{{l_{{1},{1}}}}}}=1$ when $l = {l_p} + {l_{{\tau _1}}^{\mathrm{UI}}} + {l_{{\tau _1}}^{\mathrm{IB}}}$. Moreover, the stationary positions of both the BS and the IRS imply that $\nu_1^{\mathrm{IB}} = 0$, thus avoiding the separate estimation of the Doppler shifts for $\nu_1^{\mathrm{IB}}$ and $\nu_1^{\mathrm{UI}}$. As a result, the received pilot symbols at the LoS path can be reformulated as
\begin{equation}
\begin{aligned}
&{\bf{Y}}^{\mathrm{DD}}_{\rm{p'}}\left[ {k,{l_p} + {l_{{\tau _1}}^{\mathrm{UI}}} + {l_{{\tau _1}}^{\mathrm{IB}}}} \right]  = \frac{{{x_p}}}{N}\times\\
&\sum\limits_{n=0}^{N - 1} { {{e^{\frac{{j2\pi {k_p}n}}{N}}}h_{{1},{1}}^{{\rm{UIB}}}{e^{j2\pi  {v_{{1}}^{{\rm{UI}}} } T_sn}}{e^{\frac{{ - j2\pi kn}}{N}}}} } + \mathbf{N}\left[ {k,{l_p} + {l_{{\tau _1}}^{\mathrm{UI}}} + {l_{{\tau _1}}^{\mathrm{IB}}}} \right] .
\end{aligned}
\end{equation}

Let  $\mathbf{Z}_{k,l}\triangleq \left | {\bf{Y}}^{\mathrm{DD}}_{\rm{p'}}\left[ {k,{l_p} + {l_{{\tau _1}}^{\mathrm{UI}}} + {l_{{\tau _1}}^{\mathrm{IB}}}} \right] \right |  $ denote the amplitude of the received signal, and 
\begin{equation}\label{eq:eq18}
\begin{aligned}
&\mathbf{Z}_{k,l}'\triangleq\left | \frac{{{x_p}}}{N}\sum\limits_{n=0}^{N - 1} { {{e^{\frac{{j2\pi {k_p}n}}{N}}}h_{{1},{1}}^{{\rm{UIB}}}{e^{j2\pi  {\nu_{{1}}^{{\rm{UI}}} } T_sn}}{e^{\frac{{ - j2\pi kn}}{N}}}} } \right |  \\
& \qquad\qquad= \left| {h_{1,1}^{{\rm{UIB}}}\frac{{{x_p}}}{N}\frac{{\sin \left( {\pi NT_s\left( {\nu_1^{{\rm{UI}}} + \frac{{{k_p} - k}}{{NT_s}}} \right)} \right)}}{{\sin \left( {\pi T_s\left( {\nu_1^{{\rm{UI}}} + \frac{{{k_p} - k}}{{NT_s}}} \right)} \right)}}} \right|,
\end{aligned}
\end{equation}
represents the ideal case without noise.
It can be observed that $\mathbf{Z}_{k,l}', \forall k=[1,\cdots,N]$ has a single peak, identifying the peak point (PP) as ${\hat \nu_1} = \frac{{{k_1}}}{{NT_s}} - \frac{{{k_p}}}{{NT_s}} - \nu_1^{{\rm{UI}}}$. Adjacent to the peak, the side peak points (SPP), i.e, ${\hat \nu_2} = \frac{{{k_2}}}{{NT_s}} - \frac{{{k_p}}}{{NT_s}} - \nu_1^{{\rm{UI}}}={\hat \nu_1} + \frac{1}{{NT_s}}$ and ${\hat \nu_3} = \frac{{{k_3}}}{{NT_s}} - \frac{{{k_p}}}{{NT_s}} - \nu_1^{{\rm{UI}}}={\hat \nu_1} - \frac{1}{{NT_s}}$, where  $k_1$ denotes the integer part of Doppler index corresponding to PP and $k_2=k_1+1,k_3=k_1-1$ corresponding to SPP, respectively. In such a case, the Doppler of mobile user can be estimated according to the following theorem.

\begin{theorem}\label{theorem:1}
For the ideal case that the received signal is noiseless, the value of $\nu_1^{{\rm{UI}}}$ can be estimated by the ratio of the PP and SPP \cite{tao1} {, i.e.,}
\begin{equation}\label{eq:eq19}
\hat \nu_1^{{\rm{UI}}} = \frac{{{k_1} - {k_p}}}{{NT_s}} + \frac{\psi' }{{\pi T_s}} - \frac{m}{T_s},
\end{equation}
where $m$ is an integer making ${\hat \nu_1}\in\left[ -\frac{1}{2NT_s},0\right)$, if $\mathbf{Z}_{k_2,l}'>\mathbf{Z}_{k_3,l}'$, and ${\hat \nu_1}\in\left( 0,\frac{1}{2NT_s}\right]$, if $\mathbf{Z}_{k_2,l}'<\mathbf{Z}_{k_3,l}'$, and 
\begin{equation}
\tan \psi'=\begin{cases}\frac{{\sin \left( {\frac{\pi }{N}} \right)\mathbf{Z}_{k_2,l}'}}{{\mathbf{Z}_{k_1,l}' + \mathbf{Z}_{k_2,l}'\cos \left( {\frac{\pi }{N}} \right)}},&\text{if}\quad\mathbf{Z}_{k_2,l}'>\mathbf{Z}_{k_3,l}'\\
-\frac{{\sin \left( {\frac{\pi }{N}} \right)\mathbf{Z}_{k_3,l}'}}{{\mathbf{Z}_{k_1,l}' + \mathbf{Z}_{k_3,l}'\cos \left( {\frac{\pi }{N}} \right)}},&\text{if}\quad\mathbf{Z}_{k_2,l}'<\mathbf{Z}_{k_3,l}'\end{cases}.
\end{equation}
\proof
 See Appendix \ref{app:theorem:1}.
\endproof
\end{theorem}

Theorem 1 examines the noise-free scenario to derive the estimator, which is applicable to cases with noise as well, as shown in \textbf{Algorithm} \ref{algo:1}. We first extract the signals that correspond to the LoS paths among the BS, IRS and user. {Then we} find the Doppler index corresponding to the PP and SPP, and estimate the Doppler shift $\hat \nu_1^{{\rm{UI}}}$ based on the value of $\mathbf{Z}_{k_2,l}$ and $\mathbf{Z}_{k_3,l}$ by letting
\begin{equation}\label{eq:eq21}
\hat \nu_1^{{\rm{UI}}} = \frac{{{k_1} - {k_p}}}{{NT_s}} + \frac{\psi}{{\pi T_s}} - \frac{m}{T_s},
\end{equation}
where 
\begin{equation}
\tan \psi=\begin{cases}\frac{{\sin \left( {\frac{\pi }{N}} \right)\mathbf{Z}_{k_2,l}}}{{\mathbf{Z}_{k_1,l} + \mathbf{Z}_{k_2,l}\cos \left( {\frac{\pi }{N}} \right)}},&\text{if}\quad\mathbf{Z}_{k_2,l}>\mathbf{Z}_{k_3,l}\\
-\frac{{\sin \left( {\frac{\pi }{N}} \right)\mathbf{Z}_{k_3,l}}}{{\mathbf{Z}_{k_1,l} + \mathbf{Z}_{k_3,l}\cos \left( {\frac{\pi }{N}} \right)}},&\text{if}\quad\mathbf{Z}_{k_2,l}<\mathbf{Z}_{k_3,l}\end{cases}.
\end{equation}

\begin{algorithm}[htb]
		\caption{The proposed ratio-based sensing algorithm.}\label{algo:1}
\begin{algorithmic}[1]		 			
		\State \textbf{Input} the received signals ${\bf{Y}}^{\mathrm{DD}}_{\rm{p}}\left[ {k,l} \right]$ and the delay index with the minimum delay.
		\State \qquad  Extract the signals corresponding to the LoS paths among the BS, IRS and user, i.e., ${\bf{Y}}^{\mathrm{DD}}_{\rm{p'}}\left[ {k,{l_p} + {l_{{\tau _1}}^{\mathrm{UI}}} + {l_{{\tau _1}}^{\mathrm{IB}}}} \right] $.
		\State \qquad  Find the Doppler index corresponding to the PP, i.e., $k_1$ and SPP, i.e., $k_2$ and $k_3$.
		\State \qquad Compute $\hat \nu_1^{{\rm{UI}}}$ using Eq. (\ref{eq:eq21}).
		\State  \textbf{Output} the estimation of the Doppler shift $\hat \nu_1^{{\rm{UI}}}$.
\end{algorithmic}
\end{algorithm}
\begin{remark}
The proposed estimation algorithm operates in the DD domain, capitalizing on the amplitude of received signals when $l = {l_p} + {l_{{\tau _1}}^{\mathrm{UI}}} + {l_{{\tau _1}}^{\mathrm{IB}}}$. This involves employing peak search strategies like the bisection method in the delay domain, culminating in a complexity of $\mathcal{O} \left( {{{\log }_2}\left( M \right)} \right)$.  Notably, the computational complexity of our algorithm remains unaffected by variables  ${k_i},{k_p},N,{T_s}$, and $\psi $,  maintaining an overall complexity of $\mathcal{O} \left( {{{\log }_2}\left( M \right)} \right)$.

Besides, it should be noted that the effectiveness of the proposed ratio-based sensing algorithm hinges significantly on the appropriate identification of the Doppler index corresponding to the largest and the second-largest values, as outlined in step 3 of \textbf{Algorithm} \ref{algo:1}, which is influenced by noise factors. Hence, accurate selection of the correct Doppler indexes is essential for achieving effective sensing in high-mobility systems. The performance of the proposed ratio-based sensing algorithm is given in the next subsection.

\end{remark}

\subsection{{Performance of the Proposed Sensing Algorithm}}
We first evaluate the performance of the proposed sensing algorithm from the perspective of effective sensing probability. The effective sensing probability denotes the probability of correctly selecting the Doppler indexes, i.e., PP and SPP, which is defined as Eq. (\ref{eq:eqtp}) on the top of this page. For the performance of effective sensing probability, we have the following theorem.
\begin{figure*}
\begin{equation}\label{eq:eqtp}
\mathrm{P}_{\mathrm{eff}}=
\begin{cases}\mathrm{Pr}\left[\left(\mathbf{Z}_{k_1,l}\geq\max_{k}\mathbf{Z}_{k,l}\right)\cap\left(\mathbf{Z}_{k_2,l}\geq \mathbf{Z}_{k_3,l}\right)\right],&\text{if}\quad\mathbf{Z}_{k_2,l}'>\mathbf{Z}_{k_3,l}'\\
\mathrm{Pr}\left[\left(\mathbf{Z}_{k_1,l}\geq\max_{k}\mathbf{Z}_{k,l}\right)\cap\left(\mathbf{Z}_{k_3,l}\geq \mathbf{Z}_{k_2,l}\right)\right],&\text{if}\quad\mathbf{Z}_{k_2,l}'<\mathbf{Z}_{k_3,l}'\end{cases}.
\end{equation}\hrulefill
\end{figure*}

\begin{theorem}\label{theorem:2}
In the high SNR region, the effective sensing probability can be approximated in closed form as Eq. (\ref{eq:eq22}) on the top of this page,
\begin{figure*}
\begin{equation}\label{eq:eq22}
\begin{aligned}
\mathrm{P}_{\mathrm{eff}}&\approx
\begin{cases}\frac{{\Gamma ({\varpi _3} + {\varpi _2})}}{{\Gamma ({\varpi _3} + 1)\Gamma ({\varpi _2})}}{\left( {\frac{{{\vartheta _3}}}{{{\vartheta _2}}}} \right)^{{\varpi _3}}}{}_2{F_1}\left( {{\varpi _2} + {\varpi _3},{\varpi _3};{\varpi _3} + 1; - \frac{{{\vartheta _3}}}{{{\vartheta _2}}}} \right),&\text{if}\quad\mathbf{Z}_{k_2,l}'>\mathbf{Z}_{k_3,l}'\\
\frac{{\Gamma ({\varpi _2} + {\varpi _3})}}{{\Gamma ({\varpi _2} + 1)\Gamma ({\varpi _3})}}{\left( {\frac{{{\vartheta _2}}}{{{\vartheta _3}}}} \right)^{{\varpi _2}}}{}_2{F_1}\left( {{\varpi _3} + {\varpi _2},{\varpi _2};{\varpi _2} + 1; - \frac{{{\vartheta _2}}}{{{\vartheta _3}}}} \right),&\text{if}\quad\mathbf{Z}_{k_2,l}'<\mathbf{Z}_{k_3,l}'\end{cases},
\end{aligned}
\end{equation}\hrulefill
\end{figure*}
where $\varpi_i  =  \frac{{{{\left[ {{\left(\mathbf{Z}_{k_i,l}'\right)^2} + {\sigma ^2}} \right]}^2}}}{{2{\left(\mathbf{Z}_{k_i,l}'\right)^2}{\sigma ^2} + {\sigma ^4}}}$, $\Omega_i  =  {\left(\mathbf{Z}_{k_i,l}'\right)^2} + {\sigma ^2}$, $\vartheta_i  = \frac{\varpi_i }{\Omega_i }$, $\Gamma(x)$ denotes the gamma function and ${}_2{F_1}(a, b; c;-x)$ denotes the hypergeometric function \cite{hyper}.

\proof
 See Appendix \ref{app:theorem:2}.
\endproof
\end{theorem}
\begin{remark}
The closed form of effective sensing probability serves as a key performance metric for our proposed ratio-based sensing algorithm. However, this formulation involves complex functions, including the hypergeometric and Gamma functions, which do not immediately offer intuitive insights. Therefore, we also derive achievable MSE to provide a more comprehensible understanding of its performance, which is given by the following corollary.
\end{remark}

\begin{corollary}\label{corollary:1}
Define the achievable MSE as ${\varepsilon } \buildrel \Delta \over = \left[ {|\nu _1^{{\rm{UI}}} - \hat \nu _1^{{\rm{UI}}}{|^2}} \right]$, assume that the PP and SPP can be selected successfully, {and let} $\mathbf{Z}_{k_1,l}'$ and $\mathbf{Z}_{k_2,l}'$ denote the largest and the second largest value of $\mathbf{Z}_{k,l}'$, the achievable MSE can be approximated as follows
\begin{equation}\label{eq:eq25}
\begin{aligned}
&{\varepsilon} \approx \frac{{{{\sin }^4}\psi' }}{{{{\sin }^2}\left( {\frac{\pi }{N}} \right){\pi ^2}{T_s^2}}} \times\\
&\left\{ {{{ {\frac{\mathbf{Z}_{k_1,l}'^2}{\mathbf{Z}_{k_2,l}'^2}} }} - 2 {\frac{\mathbf{Z}_{k_1,l}'}{\mathbf{Z}_{k_2,l}'}}\frac{{\Gamma ({\varpi _2} - \frac{1}{2})\Gamma ({\varpi _1} + \frac{1}{2})}}{{\Gamma ({\varpi _2})\Gamma ({\varpi _1})}}\sqrt {\frac{{{\vartheta _2}}}{{{\vartheta _1}}}}  + \frac{{{\varpi _1}}}{{{\varpi _2} - 1}}\frac{{{\vartheta _2}}}{{{\vartheta _1}}}} \right\}.
\end{aligned}
\end{equation}

Besides, an upper bound of the achievable MSE is given by
\begin{equation}\label{eq:eq26}
 {\varepsilon ^{{\rm{up}}}}=\frac{{{{\sin }^4}\psi' }}{{{{\sin }^2}\left( {\frac{\pi }{N}} \right){\pi ^2}{T_s^2}}}\left\{ {{\frac{\mathbf{Z}_{k_1,l}'^2}{\mathbf{Z}_{k_2,l}'^2}}  - 2 {\frac{\mathbf{Z}_{k_1,l}'}{\mathbf{Z}_{k_2,l}'}} \sqrt {\frac{{{\varpi _1}{\vartheta _2}}}{{{\varpi _2}{\vartheta _1}}}}  + \frac{{{\varpi _1}}}{{{\varpi _2} - 1}}\frac{{{\vartheta _2}}}{{{\vartheta _1}}}} \right\}.
\end{equation}

\proof
 See Appendix \ref{app:corollary:1}.
\endproof

The error bound of the proposed upper bound of the achievable MSE can be derived as 
\begin{equation}
\delta_{\rm{lower}} < {\varepsilon ^{{\rm{up}}}} - \varepsilon <\delta_{\rm{upper}},\end{equation}
where the lower error bound $\delta_{\rm{lower}}$ and upper error bound $\delta_{\rm{upper}}$ are respectively defined as
\begin{equation}
\begin{aligned}
&\delta_{\rm{lower}}=\frac{{2{{\sin }^4}\psi '}}{{{{\sin }^2}\left( {\frac{\pi }{N}} \right){\pi ^2}T_s^2}}\frac{{{{\bf{Z}}^{\prime} _{{k_1},{l}}}}}{{{{\bf{Z}}^{\prime} _{{k_2},{l}}}}}\sqrt {\frac{{{\varpi _1}}}{{{\varpi _2}}}} \left[ {\alpha_{\rm{lower}}  - \sqrt {\frac{{{\vartheta _2}}}{{{\vartheta _1}}}} } \right] ,\\
& \delta_{\rm{upper}}=\frac{{2{{\sin }^4}\psi '}}{{{{\sin }^2}\left( {\frac{\pi }{N}} \right){\pi ^2}T_s^2}}\frac{{{{\bf{Z}}^{\prime}_{{k_1},{l }}}}}{{{{\bf{Z}}^{\prime}_{{k_2},{l }}}}}\sqrt {\frac{{{\varpi _1}}}{{{\varpi _2}}}} \left[ {\alpha_{\rm{upper}}  - \sqrt {\frac{{{\vartheta _2}}}{{{\vartheta _1}}}} } \right],
\end{aligned}
\end{equation}
where 
\begin{equation}
\alpha_{\rm{lower}}=\frac{{{{\left( {1 + \frac{1}{{2{\varpi _1}}}} \right)}^{{\varpi _1}}}}}{{\left( {1 + \frac{1}{{12{\varpi _1}}} + \frac{1}{{288{\varpi _1}}}} \right)}}\frac{{{{\left( {1 - \frac{1}{{2{\varpi _2}}}} \right)}^{ - 1 + {\varpi _2}}}}}{{\left( {1 + \frac{1}{{12{\varpi _2}}} + \frac{1}{{288{\varpi _2}}}} \right)}},
\end{equation}
 and 
\begin{equation}\small
\begin{aligned}
\alpha_{\rm{upper}}=&{\left( {1 - \frac{1}{{2{\varpi _2}}}} \right)^{ - 1 + {\varpi _2}}}{\left( {1 + \frac{1}{{2{\varpi _1}}}} \right)^{{\varpi _1}}}\\
&\left( {1 + \frac{1}{{12\left( {{\varpi _1} + \frac{1}{2}} \right)}} + \frac{1}{{288\left( {{\varpi _1} + \frac{1}{2}} \right)}}} \right)\\
&\left( {1 + \frac{1}{{12\left( {{\varpi _2} - \frac{1}{2}} \right)}} + \frac{1}{{288\left( {{\varpi _2} - \frac{1}{2}} \right)}}} \right).
\end{aligned}
\end{equation}

\proof
 See Appendix \ref{app:Tightness}.
\endproof
\end{corollary}
\begin{remark}\label{re:remark:3}
Corollary \ref{corollary:1} derives the closed form of achievable MSE. Besides, to give further insight, we derive the upper bound of the achievable MSE by removing the complex gamma function, and the tightness of the proposed upper bound. It can be observed that the estimation accuracy mainly depends on the power of the received signal and the ratio of ${\mathbf{Z}_{k_1,l}'}$ and ${\mathbf{Z}_{k_2,l}'}$. Moreover, the estimation accuracy increases with $N$, which is due to the fact that an {increment} in $N$ enhances the resolution in the Doppler domain, thus improving the estimation accuracy.
\end{remark}

\section{{Optimization of Integrated Sensing and Communication}}\label{se:se4}
In this section, we attempt to optimize the phase shift matrix of IRS and the receive combining vector at the BS to improve the overall performance of sensing and communication in high-mobility scenarios. Prior to the optimization, we first derive the achievable rate of communication. 

Recall Eq. (\ref{eq:eq11}) and define ${\left[ {{{\bf{\Psi }}^{p_1,p_2}}} \right]_{kM + l,k'M + l'}} = {\boldsymbol{\Psi }}_{k,{k^\prime }}^{p_1,p_2}[l,l']$, the block-wise received signal can be vectorized as the following form
\begin{equation}
\begin{aligned}
{\bf{y}} & = \mathrm{vec}\left( {\mathbf{Y}^{\mathrm{DD}}}\left[ {k,l} \right]  \right)\\
& = \sum\limits_{{p_1} = 1}^{{L_{{\rm{UI}}}}}\sum\limits_{{p_2} = 1}^{{L_{{\rm{IB}}}}} { {h_{p_2}^{{\rm{IB}}}} h_{p_1}^{{\rm{UI}}}{e^{\jmath 2\pi v_{p_2}^{{\rm{IB}}}\tau _{p_1}^{{\rm{UI}}}}}} {{\bf{r}}^H}{{\bf{B}}^{{p_1},{p_2}}}{\boldsymbol{\xi }}{{\bf{\Psi }}^{{p_1},{p_2}}}{\bf{x}} + {\bf{n}},
\end{aligned}
\end{equation}
where ${{\bf{B}}^{{p_1},{p_2}}}={{{\bf{a}}_{\rm{B}}}\left( {\vartheta _{p_2}^{{\rm{IB}}}} \right){\bf{a}}_{\rm{I}}^H\left( {\phi _{p_2}^{{\rm{IB}}},\psi _{p_2}^{{\rm{IB}}}} \right)\mathrm{diag}\left( {{{\bf{a}}_{\rm{I}}}\left( {\phi _{p_1}^{{\rm{UI}}},\psi _{p_1}^{{\rm{UI}}}} \right)} \right)}$, ${\bf{x}}=\mathrm{vec}\left( {\mathbf{X}^{\mathrm{DD}}[k,l]} \right)$, and ${\bf{n}}=\mathrm{vec}\left( \mathbf{N}\left[ {k,l} \right]\right)$.

Therefore, the achievable {rate of communication} can be calculated as
\begin{equation}\label{eq:eq28}
R = \frac{1}{{MN}}{\log _2}\left| {{\bf{I}} + \gamma {{\bf{H}}_{\mathrm{truc}}}{\bf{H}}_{\mathrm{truc}}^H} \right|,
\end{equation}
where ${{\bf{H}}_{\mathrm{truc}}} \overset{\Delta}{=}   {\sum\limits_{{p_1} = 1}^{{L_{{\rm{UI}}}}}\sum\limits_{{p_2} = 1}^{{L_{{\rm{IB}}}}}  {h_{p_2}^{{\rm{IB}}}} h_{p_1}^{{\rm{UI}}}{e^{\jmath 2\pi v_{p_2}^{{\rm{IB}}}\tau _{p_1}^{{\rm{UI}}}}}} {{\bf{r}}^H}{{\bf{B}}^{{p_1},{p_2}}}{\boldsymbol{\xi }}{\bf{\Psi }}_{\mathrm{truc}}^{{p_1},{p_2}}$, $\gamma=\frac{1}{\sigma^2}$, and 
\begin{equation}
\begin{aligned}
& {\left[ {{\bf{\Psi }}_{\mathrm{truc}}^{{p_1},{p_2}}} \right]_{kM + l,k'M + l'}} \overset{\Delta}{=} \\
& \qquad\qquad\left\{ {\begin{array}{*{20}{l}}
0&{{\rm{,  if    }}\quad l' \in {{\cal L}_{g}\cup l_p},k' \in {{\cal K}_{g}}\cup k_p}\\
{{\bf{\Psi }}_{k,{k^\prime }}^{{p_1},{p_2}}\left[ {l,{l^\prime }} \right]}&{,{\rm{  {otherwise.}}}}
\end{array}} \right.
\end{aligned}.
\end{equation}

Hence, we reveal the impact of the phase shift matrix and receive combining vectors on the achievable rate of communication.

\subsection{Problem Formulation}
With the achievable MSE of sensing in Eq. (\ref{eq:eq25}) and the achievable rate of communication in Eq. (\ref{eq:eq28}), we expect to design the combining vector $\bf{r}$ and the IRS phase shifts $\boldsymbol{\xi }$, so as to maximize the achievable rate of communication subject to the constraint on the achievable MSE of sensing, which can be formulated as the following optimization problem
\begin{subequations}
\begin{flalign}
&{}&{\text{(P1)}}\quad {\mathop {\max }\limits_{{\bf{r}},{\boldsymbol{\xi }}} }&\quad R, \label{eq:eq30a}& \\
&{}&\:\,{\mathrm{s.t.}}&\quad{\varepsilon \le \gamma_1, }\label{eq:eq30b}&\\
&{}&{}&\quad {\left\| {\bf{r}} \right\|^2} = 1,\label{eq:eq30c}&\\
&{}&{}&\quad \left| {{{\left[ {\boldsymbol{\xi }} \right]}_i}} \right| = 1, \forall i\in[1,N_I]. \label{eq:eq30d}&
\end{flalign}
\end{subequations}
where $\gamma_1$ denotes the required sensing accuracy.

Solving problem (P1) presents a significant challenge due to the non-convexity nature of both the objective function and constraints. To make the above problem more tractable, we first derive the upper bound of the objective function. Following works  \cite{mimo1} and  \cite{MIMO-LB}, the above objective function can be approximated as
\begin{equation}\label{eq:equp}
\begin{aligned}
\frac{1}{{MN}}{\log _2}\left| {{\bf{I}} + \frac{\gamma}{N_s} {{\bf{H}}_{\mathrm{truc}}}{\bf{H}}_{\mathrm{truc}}^H} \right|
&\overset{(a)} \approx \frac{1}{{MN}}\sum\limits_{i=1}^{N_s} { {\log _2}\left( {1 + \gamma {{\left[ \mathbf{A} \right]}_{i,i}}} \right)}\\
&\overset{(b)} \geq  \frac{1}{{MN}}{\log _2}\left| {1 + \frac{\gamma}{N_s} \left\| {{{\bf{H}}_{\mathrm{truc}}}} \right\|_F^2} \right|,
\end{aligned}
\end{equation}
where $(a)$ is obtained {by using} the truncated singular value decomposition (SVD) of the effective truncated channel ${\bf{H}}_{\mathrm{truc}}$, $(b)$ holds according to Proposition 1 in \cite{MIMO-LB}. $\mathbf{A}$ is a diagonal matrix with the diagonal elements ${\left[ \mathbf{A} \right]}_{i,i}$ being the singular values of ${\bf{H}}_{\mathrm{truc}}$, and $N_s=N(M-2l_\mathrm{max}-1)$\footnote{Since  Eq. (\ref{eq:equp}) provides an approximate representation of the objective function for  (P1), the solution is suboptimal.}.
Moreover, recall Eq. (\ref{eq:eq18}) and define $a_{k_i}=\frac{{{x_p}}}{N}\frac{{\sin \left( {\pi NT_s\left( {\nu_1^{{\rm{UI}}} + \frac{{{k_p} - k_i}}{{NT_s}}} \right)} \right)}}{{\sin \left( {\pi T_s\left( {\nu_1^{{\rm{UI}}} + \frac{{{k_p} - k_i}}{{NT_s}}} \right)} \right)}}$, $\mathbf{Z}_{k,l}'$ can be further rewritten as a more compact form $\mathbf{Z}_{k_i,l}'=  \left| {h_{1,1}^{{\rm{UIB}}}a_{k_i}} \right|$, and $\mathrm{tan} \psi' =\frac{{\sin \left( {\frac{\pi }{N}} \right)a_{k_2}}}{{a_{k_1} + a_{k_2}\cos \left( {\frac{\pi }{N}} \right)}}$. Then, according to Eq. (\ref{eq:eq26}), the upper bound of the achievable MSE can be further rewritten as Eq. (\ref{eq:eqeup}) on the top of this page.
\begin{figure*}
\begin{equation}\label{eq:eqeup}
\varepsilon^{\mathrm{up}}=\frac{{{{\sin }^4}\psi' }}{{{{\sin }^2}\left( {\frac{\pi }{N}} \right){\pi ^2}{T_s^2}}}\left\{ {\frac{\left|a_{{k_1}}\right|^2}{\left|{a_{{k_2}}}\right|^2}  - 2 \frac{\left|a_{{k_1}}\right|}{\left|{a_{{k_2}}}\right|}\sqrt {\frac{{{{\left| {h_{1,1}^{{\rm{UIB}}}{a_{{k_1}}}} \right|}^2} + {\sigma ^2}}}{{{{\left| {h_{1,1}^{{\rm{UIB}}}{a_{{k_2}}}} \right|}^2} + {\sigma ^2}}}}   + \frac{{\left( {{{\left| {h_{1,1}^{{\rm{UIB}}}{a_{{k_1}}}} \right|}^2} + {\sigma ^2}} \right)\left( {{{\left| {h_{1,1}^{{\rm{UIB}}}{a_{{k_2}}}} \right|}^2} + {\sigma ^2}} \right)}}{{{{\left| {h_{1,1}^{{\rm{UIB}}}{a_{{k_2}}}} \right|}^4}}}} \right\}.
\end{equation} 
\hrulefill
\end{figure*}
It can be verified that the upper bound of achievable MSE is a decreasing function with respect to $\left| {h_{1,1}^{{\rm{UIB}}}}\right|$   by using the first-order derivation. Therefore, the constraint function, i.e., Eq. (\ref{eq:eq30b}), is equivalent to $\left| {h_{1,1}^{{\rm{UIB}}}}\right|\ge \gamma'$, where $\gamma'$ can be obtained by solving the equation $\varepsilon^{\mathrm{up}} = \gamma_1 $. Hence, the problem (P1) can be rewritten as
\begin{subequations}
\begin{flalign}
&{}&{\text{(P2)}}\quad {\mathop {\max }\limits_{{\bf{r}},{\boldsymbol{\xi }}} }&\quad {\left\| {{{\bf{H}}_{\mathrm{truc}}}} \right\|_F^2}, \label{eq:eq33a}& \\
&{}&\:\,{\mathrm{s.t.}}&\quad{\left| {h_{1,1}^{{\rm{UIB}}}}\right|\ge \gamma', }\label{eq:eq33b}&\\
&{}&{}&\quad {\left\| {\bf{r}} \right\|^2} = 1,\label{eq:eq33c}&
\\&{}&{}&\quad \left| {{{\left[ {\boldsymbol{\xi }} \right]}_i}} \right| = 1, \forall i\in[1,N_I].\label{eq:eq33d}&
\end{flalign}
\end{subequations}

The problem (P2) is not a jointly convex problem of $\mathbf{r}$ and $\boldsymbol{\xi}$. In this context, we employ an alternative optimization method to solve it. Specifically, we divide it into two subproblems. One optimizes $\mathbf{r}$ with a given $\boldsymbol{\xi}$, the other optimizes $\boldsymbol{\xi}$ with a given $\mathbf{r}$. The two subproblems are iteratively optimized until convergence.
\subsection{{Algorithm Design}}
For a given $\boldsymbol{\xi}$, the objective function of (P2) can be rewritten as
\begin{equation}
\begin{aligned}
\left\| {{{\bf{H}}_{\mathrm{truc}}}} \right\|_F^2 &= {\mathrm{Tr}}\left( {{\bf{H}}_{\mathrm{truc}}^H{{\bf{H}}_{\mathrm{truc}}}} \right)\\
&=\sum\limits_{{p_{2}} = 1}^{{L_{{\rm{IB}}}}} \sum\limits_{{p_{1}} = 1}^{{L_{{\rm{UI}}}}} \sum\limits_{{p_{2}'} = 1}^{{L_{{\rm{IB}}}}} \sum\limits_{{p_{1}'} = 1}^{{L_{{\rm{UI}}}}} {\alpha _{{p_{1}},{p_{2}},{p_{1}'},{p_{2}'}}}{{\bf{r}}^H}{{\bf{a}}_{\rm{B}}}\left( {\vartheta _{p_{2}}^{{\rm{IB}}}} \right)\\
& \qquad\quad \times {\bf{a}}_{\rm{B}}^H\left( {\vartheta _{{p_{2}'}}^{{\rm{IB}}}} \right){\bf{r}}{\mathrm{Tr}}\left\{ {{\bf{\Psi }}_{\mathrm{truc}}^{{p_{1}},{p_{2}}}{{\left( {{\bf{\Psi }}_{\mathrm{truc}}^{{p_{1}'},{p_{2}'}}} \right)}^H}} \right\}    \\
& = {{\bf{r}}^H}{\bf{Ar}},
\end{aligned}
\end{equation}
where 
\begin{equation}
\begin{aligned}
&\mathbf{A}=\sum\limits_{{p_2} = 1}^{{L_{{\rm{IB}}}}} \sum\limits_{{p_1} = 1}^{{L_{{\rm{UI}}}}} \sum\limits_{{p'_2} = 1}^{{L_{{\rm{IB}}}}} \sum\limits_{{p'_1} = 1}^{{L_{{\rm{UI}}}}} {\alpha _{{p_1},{p_2},{p'_1},{p'_2}}}{{\bf{a}}_{\rm{B}}}\left( {\vartheta _{p_2}^{{\rm{IB}}}} \right)\times \\
&\qquad\qquad\quad{\bf{a}}_{\rm{B}}^H\left( {\vartheta _{{p'_2}}^{{\rm{IB}}}} \right){\mathrm{Tr}}\left\{ {{\bf{\Psi }}_{\mathrm{truc}}^{{p_1},{p_2}}{{\left( {{\bf{\Psi }}_{\mathrm{truc}}^{{p'_1},{p'_2}}} \right)}^H}} \right\},
\end{aligned}
\end{equation}
 and 
\begin{equation}\label{eq:eqH}
\begin{aligned}
{{\alpha _{{p_1},{p_2},{p'_1},{p'_2}}}}
&=h_{p_2}^{{\rm{IB}}}h_{p_1}^{{\rm{UI}}}{{\left( {h_{{p'_2}}^{{\rm{IB}}}h_{{p'_1}}^{{\rm{UI}}}} \right)}^*}{e^{\jmath 2\pi v_{p_2}^{{\rm{IB}}}\tau _{p_1}^{{\rm{UI}}} - \jmath 2\pi v_{{p'_2}}^{{\rm{IB}}}\tau _{{p'_1}}^{{\rm{UI}}}}}\times\\
&\qquad{\bf{a}}_{\rm{I}}^H\left( {\phi _{p_2}^{{\rm{IB}}},\psi _{p_2}^{{\rm{IB}}}} \right)\mathrm{{\mathrm{diag}}}\left( {{{\bf{a}}_{\rm{I}}}\left( {\phi _{p_1}^{{\rm{UI}}},\psi _{p_1}^{{\rm{UI}}}} \right)} \right){\boldsymbol{\xi }}\times\\
&\qquad{{\boldsymbol{\xi }}^H}\mathrm{{\mathrm{diag}}}\left( {{\bf{a}}_{\rm{I}}^H\left( {\phi _{{p'_1}}^{{\rm{UI}}},\psi _{{p'_1}}
^{{\rm{UI}}}} \right)} \right){\bf{a}}_{\rm{I}}^{}\left( {\phi _{{p'_2}}^{{\rm{IB}}},\psi _{{p'_2}}^{{\rm{IB}}}} \right).
\end{aligned}
\end{equation}

Thus, the problem (P2) can be simplified to the subproblem (P3) by several mathematical manipulations\footnote{Given that both sides of Eq. (\ref{eq:eq33b}) represent positive values, squaring them simplifies the derivation of subspace methods, leading to the formulation of the constraint in Eq. (\ref{eq:eq34b}).}
\begin{subequations}
\begin{flalign}
&{}&{\text{(P3)}}\quad {\mathop {\max }\limits_{{\bf{r}}} }&\quad {{{\bf{r}}^H}{\bf{Ar}}}, \label{eq:eq34a}& \\
&{}&\:\,{\mathrm{s.t.}}&\quad{\left | {{\bf{r}}^H}{{\bf{a}}_{\rm{B}}}\left( {\vartheta _{{1}}^{{\rm{IB}}}} \right) \right | ^2 \ge \lambda_{\mathrm{r}} },\label{eq:eq34b},&\\
&{}&{}&\quad {\left\| {\bf{r}} \right\|^2} = 1.\label{eq:eq34c}&
\end{flalign}
\end{subequations}
where 
\begin{equation}
\lambda_{\mathrm{r}}=\left({\frac{{\gamma '}}{{\left| {h_1^{{\rm{IB}}}h_1^{{\rm{UI}}}} \right|\left| {{\bf{a}}_{\rm{I}}^H\left( {\phi _{{1}}^{{\rm{IB}}},\psi _{{1}}^{{\rm{IB}}}} \right)\boldsymbol{\Theta} {{\bf{a}}_{\rm{I}}}\left( {\phi _{1}^{{\rm{UI}}},\psi _{1}^{{\rm{UI}}}} \right)} \right|}}}\right)^2 ,
\end{equation}

The above problem is a quadratically constrained quadratic programming (QCQP) problem, which can be addressed using the semidefinite relaxation (SDR) method. However, this method suffers from substantial computational complexity.

It is noted that the solution of $\mathbf{r}$ always satisfies that ${\bf{r}} = a{{\bf{u}}_{\bf{\alpha }}} + b{{\bf{u}}_{\bf{\beta }}}$, where $\left\| {{{\bf{u}}_{\bf{\alpha }}}} \right\| = \left\| {{{\bf{u}}_{\bf{\beta }}}} \right\| = 1$,  ${\bf{u}_{\bf{\alpha }}} \in \mathrm{span}\left\{ {{{\bf{u}}_1},{{\bf{a}}_{\rm{B}}}\left( {\vartheta _{{1}}^{{\rm{IB}}}} \right)} \right\}$, and ${\bf{u}_{\bf{\beta }}} \bot \mathrm{span}\left\{ {{{\bf{u}}_1},{{\bf{a}}_{\rm{B}}}\left( {\vartheta _{{1}}^{{\rm{IB}}}} \right)} \right\}$. The constraint can be rewritten as ${\left\| {\bf{r}} \right\|^2} = {a^2} + {b^2} \le 1$ and ${\left | {{\bf{r}}^H}{{\bf{a}}_{\rm{B}}}\left( {\vartheta _{{1}}^{{\rm{IB}}}} \right) \right | ^2={a^2}{\left\| {{\bf{a}}_{\rm{B}}^H\left( {\vartheta _{{1}}^{{\rm{IB}}}} \right){{\bf{u}}_{\bf{\alpha }}}} \right\|^2}} \ge \lambda_{\mathrm{r}}$
 \cite{subspace1}. {As a special case where $a=1$ {and} $b=0$, the solution to the problem {is} ${\bf{r}} = {{\bf{u}}_{\bf{\alpha }}}$.
Additionally, it is imperative to fully utilize the power constraint to optimize the objective function, as noted in \cite{subspace2}. Consequently, when the constraint of Eq. (\ref{eq:eq34b}) is met, the problem aligns with the Rayleigh entropy problem, where the optimal solution can be expressed as $\mathbf{r}^*=\mathbf{u}_1$, with ${{\bf{u}}_1}$ representing the eigenvector associated with the largest eigenvalue of matrix $\mathbf{A}$.} Otherwise,  ${\bf{r}}^* = {{\bf{u}}_{\bf{\alpha }}}= {x_1}{\boldsymbol{\alpha }} + {x_2}{\boldsymbol{\beta }}$, where $\boldsymbol{\alpha }$ and $\boldsymbol{\beta }$ can be obtained via the Gram-Schmidt orthogonalization method and is given by ${\boldsymbol{\alpha }} = \frac{{{\bf{a}}_{\rm{B}}^H\left( {\vartheta _{{p_2}}^{{\rm{IB}}}} \right)}}{{\left\| {{\bf{a}}_{\rm{B}}^H\left( {\vartheta _{{p_2}}^{{\rm{IB}}}} \right)} \right\|}},{\boldsymbol{\beta }} = \frac{{{{\bf{u}}_1} - \left( {{{\bf{\alpha }}^H}{{\bf{u}}_1}} \right){\bf{\alpha }}}}{{\left\| {{{\bf{u}}_1} - \left( {{{\bf{\alpha }}^H}{{\bf{u}}_1}} \right){\bf{\alpha }}} \right\|}}$. To this end,  by substituting $r^{*}$ into the objective function of (P3), i.e., Eq. (\ref{eq:eq34a}), the optimization problem can be further rewritten as (P3-1)
\begin{subequations}
\begin{flalign}
&{}&{\text{(P3-1)}}\quad {\mathop {\max }\limits_{{x_1,x_2}} }&\quad {{{\left( {{x_1}{\boldsymbol{\alpha }} + {x_2}{\boldsymbol{\beta }}} \right)}^H}{\bf{A}}\left( {{x_1}{\boldsymbol{\alpha }} + {x_2}{\boldsymbol{\beta }}} \right)}, \label{eq:eq37a}& \\
&{}&\:\,{\mathrm{s.t.}}&\quad{{{{\left| {{x_1}} \right|}^2}{{\left\| {{\bf{a}}_{\rm{B}}^H\left( {\vartheta _{{1}}^{{\rm{IB}}}} \right)} \right\|}^2}} \ge\lambda_{\mathrm{r}} }\label{eq:eq37b},&\\
&{}&{}&\quad {{{\left| {{x_1}} \right|}^2} + {{\left| {{x_2}} \right|}^2} = 1}.\label{eq:eq37c}&
\end{flalign}
\end{subequations}

According to the Karush-Kuhn-Tucker (KKT) condition, the closed form of the optimal solution is given by
\begin{equation}
{x_1} = \sqrt {\frac{\lambda_{\mathrm{r}} }{{{{\left\| {{\bf{a}}_{\rm{B}}^H\left( {\vartheta _{{p_2}}^{{\rm{IB}}}} \right)} \right\|}^2}}}} {\bar x_1},{x_2} = \sqrt {1 - \frac{\lambda_{\mathrm{r}} }{{{{\left\| {{\bf{a}}_{\rm{B}}^H\left( {\vartheta _{{p_2}}^{{\rm{IB}}}} \right)} \right\|}^2}}}} {\bar x_2},
\end{equation}
where ${\bar x_1}$ and ${\bar x_2}$ denote the direction of $x_1$ and $x_2$, respectively. Since the optimal direction aligns with solving the Rayleigh entropy problem, the problem (P3-1) {is equivalent to} finding the subspace that most closely correlates with subspace $\mathbf{u}_1$, essentially aiming to maximize ${\left| {{{\left( {{x_1}{\boldsymbol{\alpha }} + {x_2}{\boldsymbol{\beta }}} \right)}^H}{{\bf{u}}_1}} \right|}$. It becomes evident that the optimal direction of $x_1$ and $x_2$ can be derived as ${\bar x_1} = \frac{{{\bf{u}}_1^H{\boldsymbol{\alpha }}}}{{\left| {{\bf{u}}_1^H{\boldsymbol{\alpha }}} \right|}},{\bar x_2} = \frac{{{\bf{u}}_1^H{\boldsymbol{\beta }}}}{{\left| {{\bf{u}}_1^H{\boldsymbol{\beta }}} \right|}}$.

Therefore, the optimal solution can be derived as 
\begin{equation}\label{eq:eq39}
{{\bf{r}}^*} =\begin{cases}\mathbf{u}_1,&\text{if}\quad {\left | {\mathbf{u}_1^H}{{\bf{a}}_{\rm{B}}}\left( {\vartheta _{{1}}^{{\rm{IB}}}} \right) \right | ^2 \ge\lambda_{\mathrm{r}} }\\
{x_1}{\boldsymbol{\alpha }} + {x_2}{\boldsymbol{\beta }},&\text{others},\end{cases},
\end{equation}
where $
{x_1} = \sqrt {\frac{\lambda_{\mathrm{r}} }{{{{\left\| {{\bf{a}}_{\rm{B}}^H\left( {\vartheta _{{1}}^{{\rm{IB}}}} \right)} \right\|}^2}}}} \frac{{{\bf{u}}_1^H{\boldsymbol{\alpha }}}}{{\left| {{\bf{u}}_1^H{\boldsymbol{\alpha }}} \right|}}$, ${x_2} = \sqrt {1 - \frac{\lambda_{\mathrm{r}} }{{{{\left\| {{\boldsymbol{a}}_{\rm{B}}^H\left( {\vartheta _{{1}}^{{\rm{IB}}}} \right)} \right\|}^2}}}} \frac{{{\bf{u}}_1^H{\boldsymbol{\beta }}}}{{\left| {{\bf{u}}_1^H{\boldsymbol{\beta }}} \right|}}$,
 and
$
{\boldsymbol{\alpha }} = \frac{{{\boldsymbol{a}}_{\rm{B}}^H\left( {\vartheta _{{1}}^{{\rm{IB}}}} \right)}}{{\left\| {{\boldsymbol{a}}_{\rm{B}}^H\left( {\vartheta _{{1}}^{{\rm{IB}}}} \right)} \right\|}}$, ${\boldsymbol{\beta }} = \frac{{{{\bf{u}}_1} - \left( {{{\boldsymbol{\alpha }}^H}{{\bf{u}}_1}} \right){\boldsymbol{\alpha }}}}{{\left\| {{{\bf{u}}_1} - \left( {{{\boldsymbol{\alpha }}^H}{{\bf{u}}_1}} \right){\boldsymbol{\alpha }}} \right\|}}$.


Similar to Eq. (\ref{eq:eqH}), with a given $\bf{r}$,  problem (P2) can be reformulated as subproblem (P4)
\begin{subequations}
\begin{flalign}
&{}&{\hspace{-3mm}\text{(P4)}} \quad{\mathop {\max }\limits_{{\boldsymbol{\xi}}} }&\quad {{{\boldsymbol{\xi}}^H}{\bf{B}}\boldsymbol{\xi}} \label{eq:eq40.1a},& \\
&{}&\:\,{\mathrm{s.t.}}&\quad{{{\left| {{{\boldsymbol{\xi }}^H}\mathrm{diag}\left( {{\bf{a}}_{\rm{I}}^H\left( {\phi _{1}^{{\rm{UI}}},\psi _{1}^{{\rm{UI}}}} \right)} \right){\bf{a}}_{\rm{I}}^{}\left( {\phi _{{1}}^{{\rm{IB}}},\psi _{{1}}^{{\rm{IB}}}} \right)} \right|}^2} \ge \lambda_{\mathrm{\xi}}}\label{eq:eq40b},&\\
&{}&{}&\quad \left| {{{\left[ {\boldsymbol{\xi }} \right]}_i}} \right| = 1, \forall i\in[1,N_I],\label{eq:eq40c}&
\end{flalign}
\end{subequations}
where 
\begin{equation}
 \lambda_{\mathrm{\xi}}={{\left( {\frac{{\gamma '}}{{\left| {h_1^{{\rm{IB}}}h_1^{{\rm{UI}}}{{\bf{r}}^H}{{\bf{a}}_{\rm{B}}}\left( {\vartheta _{{1}}^{{\rm{IB}}}} \right)} \right|}}} \right)}^2},
\end{equation}
and
\begin{equation}
\begin{aligned}
&\mathbf{B}
=\sum\limits_{{p_2} = 1}^{{L_{{\rm{IB}}}}} {\sum\limits_{{p_1} = 1}^{{L_{{\rm{UI}}}}} {\sum\limits_{{p'_2} = 1}^{{L_{{\rm{IB}}}}} {\sum\limits_{{p'_1} = 1}^{{L_{{\rm{UI}}}}}}}}
{\beta _{{p_1},{p_2},{p'_1},{p'_2}}}\mathrm{Tr}\left\{ {{\bf{\Psi }}_{\mathrm{truc}}^{{p_1},{p_2}}{{\left( {{\bf{\Psi }}_{\mathrm{truc}}^{{p'_1},{p'_2}}} \right)}^H}} \right\}\times\\
&\qquad\qquad\qquad \mathrm{diag}\left( {{\bf{a}}_{\rm{I}}^H\left( {\phi _{{p'_1}}^{{\rm{UI}}},\psi _{{p'_1}}^{{\rm{UI}}}} \right)} \right)
{\bf{a}}_{\rm{I}}^{}\left( {\phi _{{p'_2}}^{{\rm{IB}}},\psi _{{p'_2}}^{{\rm{IB}}}} \right)\times\\
&\qquad\qquad\qquad{\bf{a}}_{\rm{I}}^H\left( {\phi _{p_2}^{{\rm{IB}}},\psi _{p_2}^{{\rm{IB}}}} \right)\mathrm{diag}\left( {{{\bf{a}}_{\rm{I}}}\left( {\phi _{p_1}^{{\rm{UI}}},\psi _{p_1}^{{\rm{UI}}}} \right)} \right).
\end{aligned}
\end{equation}

Similarly, when the constraint Eq. (\ref{eq:eq40b}) is met,  the problem (P4) is equivalent to finding the subspace that is most correlated with the subspace $\mathbf{u}_2$, i.e.,
\begin{subequations}
\begin{flalign}
&{}&{\text{(P4-1)}}\quad {\mathop {\max }\limits_{{\boldsymbol{\xi}}} }&\quad {{{\left| {{{\boldsymbol{\xi }}^H}{{\bf{u}}_2 }} \right|}^2}} \label{eq:eq40a},& \\
&{}&{{\mathrm{s.t.}}}&\quad \left| {{{\left[ {\boldsymbol{\xi }} \right]}_i}} \right| = 1, \forall i\in[1,N_I],\label{eq:eq40c}&
\end{flalign}
\end{subequations}
where ${{\bf{u}}_2}$ is the eigenvector corresponding to the largest eigenvalue of matrix $\mathbf{B}$. Therefore, the optimal solution of $\boldsymbol{\xi}^*$ can be derived as 
\begin{equation}\label{eq:eq43}
\boldsymbol{\xi}^*= \exp \left( {j\angle {{\bf{u}}_2 }} \right),
\end{equation}
where $\angle {{\bf{u}}_2 }$ denotes the angle of ${{\bf{u}}_2 }$.

However, when Eq. (\ref{eq:eq40b}) is violated, it is hard to obtain the optimal solution that satisfies the unit-modulus phase-shift constraints. 

In view of this, we employ the consensus alternative direction method of multipliers (ADMM) approach \cite{CADMM} that converts the optimization problem into several QCQP with a single constraint problem (QCQP-1). This allows for parallel updating of the optimization variables. To facilitate this process, we introduce auxiliary variables $\mathbf{z}_1$ and $\mathbf{z}_2$, and the optimization problem can be rewritten as
\begin{subequations}
\begin{flalign}
&{}&{\text{(P4-2)}} {\mathop {\min }\limits_{{\boldsymbol{\xi}},\mathbf{z}_1,\mathbf{z}_2} }& -{{{\boldsymbol{\xi}}^H}{\bf{B}}\boldsymbol{\xi}}+{\rho \sum\limits_{i = 1}^2 {{{\left\| {{{\bf{z}}_i} - {\boldsymbol{\xi }} + {{\boldsymbol{\mu}}_i}} \right\|}^2}} } \label{eq:eq43a},& \\
&{}&\hspace{-3mm}{\mathrm{s.t.}}&{{{\left| {{{\mathbf{ z}}_2^H}\mathrm{diag}\left( {{\bf{a}}_{\rm{I}}^H\left( {\phi _{1}^{{\rm{UI}}},\psi _{1}^{{\rm{UI}}}} \right)} \right){\bf{a}}_{\rm{I}}^{}\left( {\phi _{{1}}^{{\rm{IB}}},\psi _{{1}}^{{\rm{IB}}}} \right)} \right|}^2} \ge \lambda_{\mathrm{\xi}}}\label{eq:eq43b},&\\
&{}&{}& \left| {{{\left[ {\mathbf{z }_1} \right]}_i}} \right| = 1, \forall i\in[1,N_I],\label{eq:eq43c}&
\end{flalign}
\end{subequations}
where $\rho$ and $\boldsymbol{\mu}_i$  are the penalty factor and the dual variable vector. Thus, the optimization problem can be solved by iteratively operating the following four steps until convergence.

\text{\bf{Step 1:}} Fixing $\mathbf{z}_i, i = {1,2}$, the problem can be reformulated as
\begin{equation}
\text{(P4-3)}\min_{\boldsymbol{\xi}}\quad-\mathbf{\boldsymbol{\xi}}^H\mathbf{B}\boldsymbol{\xi}+\rho\sum_{i=1}^2\left\|\mathbf{z}_i-\mathbf{\xi}+\boldsymbol{\mu}_i\right\|^2,
\end{equation}
which can be optimally solved by utilizing the first-order optimality condition
\begin{equation}\label{eq:eq46}
{\boldsymbol{\xi }} = {\left( {2\rho {{\bf{I}}_{{N_I}}} - {\bf{B}}} \right)^{ - 1}}\left[ {\rho \sum\limits_{i = 1}^2 {\left( {{{\bf{z}}_i} + {{\boldsymbol{\mu}}_i}} \right)} } \right].
\end{equation}

\text{\bf{Step 2:}} Fixing $\boldsymbol{\xi}$ and $\mathbf{z}_2$, the problem can be reformulated as
\begin{equation}
\begin{array}{*{20}{l}}
\text{(P4-4)}&{\mathop {\min }\limits_{{{\bf{z}}_1}} }&{\rho {{\left\| {{{\bf{z}}_1} - {\bf{\xi }} + {{\boldsymbol{\mu}}_1}} \right\|}^2}},\\
&{s.t.}&{\left| {{{\left[ {{{\bf{z}}_1}} \right]}_i}} \right| = 1}, \forall i\in[1,N_I],
\end{array}
\end{equation}
which can be optimally solved by
\begin{equation}\label{eq:eq48}
{{\bf{z}}_1} = \exp \left( {j\angle \left( {{\boldsymbol{\xi }} - {{\boldsymbol{\mu}}_1}} \right)} \right).
\end{equation}

\text{\bf{Step 3:}} Fixing $\boldsymbol{\xi}$ and $\mathbf{z}_1$, the problem can be reformulated as
\begin{equation}
\begin{array}{*{20}{l}}
\text{(P4-5)}&\hspace{-3mm}{\mathop {\min }\limits_{{{\bf{z}}_2}} }&\hspace{-3mm}{\rho {{\left\| {{{\bf{z}}_2} - {\bf{\xi }} + {{\boldsymbol{\mu}}_2}} \right\|}^2}},\\
&\hspace{-3mm}{s.t.}&\hspace{-3mm}{{{\left\| {{\bf{z}}_2^H\mathrm{diag}\left( {{\bf{a}}_{\rm{I}}^H\left( {\phi _{p1}^{{\rm{UI}}},\psi _{p1}^{{\rm{UI}}}} \right)} \right){\bf{a}}_{\rm{I}}^{}\left( {\phi _{{p_2}}^{{\rm{IB}}},\psi _{{p_2}}^{{\rm{IB}}}} \right)} \right\|}^2} \ge \lambda_{\mathrm{\xi}}},
\end{array}
\end{equation}
which can be optimally solved \cite{CADMM} by
\begin{equation}\label{eq:eq50}
{{\bf{z}}_2} = {\boldsymbol{\zeta }} + \frac{{{{\left[ {\sqrt {{\lambda_{\mathrm{\xi}} }}  - \left| {{{\boldsymbol{\zeta }}^H}{{\bf{a}}_\theta }} \right|} \right]}_ + }}}{{{{\left\| {{{\bf{a}}_\theta }} \right\|}^2}\left| {{{\boldsymbol{\zeta }}^H}{{\bf{a}}_\theta }} \right|}}{{\bf{a}}_\theta }{\bf{a}}_\theta ^H{\boldsymbol{\zeta }},
\end{equation}
where ${\boldsymbol{\zeta }} = {\boldsymbol{\xi }} - {{\boldsymbol{\mu}}_2},{{\bf{a}}_\theta } = \mathrm{diag}\left( {{\bf{a}}_{\rm{I}}^H\left( {\phi _{1}^{{\rm{UI}}},\psi _{1}^{{\rm{UI}}}} \right)} \right){\bf{a}}_{\rm{I}}^{}\left( {\phi _{{1}}^{{\rm{IB}}},\psi _{{1}}^{{\rm{IB}}}} \right)$ and $[x]_{+}=\max\{0,x\}$.

\text{\bf{Step 4:}} Update the dual variables by 
\begin{equation}\label{eq:eq51}
{{\boldsymbol{\mu}}_i} = {{\bf{z}}_i} - {\bf{\xi }} + {{\boldsymbol{\mu}}_i}.
\end{equation}

For the above iterative solving approach, it can be guaranteed to converge to a set of stationary solutions as proved in \cite{CADMM} and achieves a set of solutions that satisfy the constraint $\boldsymbol{\xi}=\mathbf{z}_i,i=1,2$.

In summary, the proposed optimization algorithm for integrated sensing and communication is given by \textbf{Algorithm} \ref{algo:2}, where $\epsilon _1$ is the estimation accuracy for optimizing $\boldsymbol{\xi}$. A lower $\rho$ value steers $\boldsymbol{\xi}$ update towards maximizing the objective function described in Eq. (\ref{eq:eq40.1a}), while a higher $\rho$ value ensures alignment with auxiliary variables $\mathbf{z}_i$, each adhering to the constraint $\mathbf{z}_i=\boldsymbol{\xi}$.
Moreover, starting with a feasible $\boldsymbol{\xi}^0$ facilitates a smaller $\rho$, promoting rapid objective reduction in Eq. (\ref{eq:eq43a}) without the risk of infeasibility \cite{CADMM}. Therefore, initializing optimization variables with a feasible solution $\boldsymbol{\xi}^0$ and $\mathbf{r}^0$  can significantly enhance the convergence speed of \textbf{Algorithm} \ref{algo:2}. For this purpose, we initialize ${{\bf{r}}^0}={{\bf{a}}_{\rm{B}}}\left( {\vartheta _1^{{\rm{IB}}}} \right)$ and ${\boldsymbol{\xi }^0}=\mathrm{diag}{{\bf{a}}_{\rm{I}}}\left( {\phi _1^{{\rm{UI}}},\psi _1^{{\rm{UI}}}} \right){\bf{a}}_{\rm{I}}^H\left( {\phi _1^{{\rm{IB}}},\psi _1^{{\rm{IB}}}} \right)$ to maximize $\left| {h_{1,1}^{{\rm{UIB}}}}\right|$, set $\mathbf{z}_i^0=\boldsymbol{\xi}^0$, ${{\boldsymbol{\mu}}_i^0}=\bf{0}$,  and choose $\rho$ as the largest eigenvalue of $\mathbf{B }$, denoted as $\lambda \left ( \mathbf{B } \right ) _{\rm{max}}$, to ensure the convexity of Eq. (\ref{eq:eq43a}).

\subsection{Algorithm Analysis}
In this subsection, we analyze the proposed subspace-based beamforming design algorithm from the perspectives of convergence behaviour and computational complexity.
\subsubsection {Convergence Behavior}
\ \newline
\indent{The upper bound of the objective function $R$ is derived in Eq. (\ref{eq:equp}), and the optimization variables are decoupled properly by dividing the original problem (P1) into (P3) and (P4). For optimizing the combining vector $\mathbf{r}$, the optimal solution is derived and the condition, i.e., $R(\mathbf{r}^t,\boldsymbol{\xi}^t)\le R(\mathbf{r}^{t+1},\boldsymbol{\xi}^t)$ can be satisfied. Similarly, for optimizing the phase shift vector $\boldsymbol{\xi}$, the solution of the consensus ADMM approach also satisfies the condition, i.e., $R(\mathbf{r}^{t+1},\boldsymbol{\xi}^t)\le R(\mathbf{r}^{t+1},\boldsymbol{\xi}^{t+1})$. Therefore, the proposed algorithm is non-decreasing and upper-bounded, thus the convergence of the proposed algorithm can be guaranteed.}

\subsubsection {Computational Complexity}
\ \newline
\indent{The overall computational complexity of the proposed beamforming design algorithm mainly depends on calculating the matrix $\mathbf{A}$,  matrix $\mathbf{B}$ and performing the matrix inversion in Eq. (\ref{eq:eq46}), which are $\mathcal{O}\left( {\left( {{N^3}{M^3} + N_B^2} \right)L_{{\rm{UI}}}^2L_{{\rm{IB}}}^2} \right)$, $\mathcal{O}\left( {\left( {{N^3} {M^3} + N_I^2} \right)L_{{\rm{UI}}}^2L_{{\rm{IB}}}^2} \right)$, and $\mathcal{O}\left( {N_I^3} \right)$, respectively.
Therefore, the overall complexity of \textbf{Algorithm} \ref{algo:2} is $\mathcal{O}\left(T_1N_I^3+ L_{{\rm{UI}}}^2L_{{\rm{IB}}}^2\left( T_1{N^3}{M^3} + T_1N_I^2+N_B^2 \right) \right)$, where $T_1$ denotes the maximum number of iterations.}
\begin{table}[ht]
\renewcommand\arraystretch{1.3}
\begin{center}
\caption{Computational complexity comparison.}
\label{tab1}
\vspace{-3mm}    
\begin{tabular}{ | m{4.1cm} <{\centering}| m{4.0cm}<{\centering} |}
    \hline
    \textbf{Algorithm} &\textbf{Computational complexity}  \\  \hline
Proposed ratio-based algorithm& $\mathcal{O} \left( {{{\log }_2}\left( M \right)} \right)$    \\ \hline
Embedded algorithm in \cite{embedded1}& 	$\mathcal{O} \left( N{{{\log }_2}\left( M \right)} \right)$   \\ \hline
MUSIC algorithm& 	$\mathcal{O}\left( {{{\log }_2}\left( M \right) + {N^3} + N{K_N}} \right)$    \\ \hline
TLS-ESPRIT algorithm in \cite{TLS-ESPRIT}& 	$\mathcal{O}\left( {{{\log }_2}\left( M \right) + {N^3}} \right)$    \\ \hline
    \end{tabular}
\end{center}
\end{table}

\begin{algorithm}[htb!]
		\caption{The proposed subspace-based beamforming design {algorithm}}\label{algo:2}
\begin{algorithmic}[1]
		\State Initialize variables $\mathbf{r}^0 $, $\mathbf{z}_i^0$, ${{\boldsymbol{\mu}}_i^0}$, $\boldsymbol{\xi}^0$, $\rho$, and set the iteration index $t=1$.
		\Repeat
			\State  Update the combining vector $\mathbf{r}^{t}$ by (\ref{eq:eq39}).
				\If {the constraint, Eq. (\ref{eq:eq40b}) is satisfied}
				\State  Update the phase shift vector $\boldsymbol{\xi}^{t}$ by (\ref{eq:eq43}).
				\Else
				\Repeat
					\State  Update the phase shift vector $\boldsymbol{\xi}$ by (\ref{eq:eq46}).
					\State  Update the auxiliary variables $\mathbf{z}_1$ and $\mathbf{z}_2$ by (\ref{eq:eq48}) and (\ref{eq:eq50}), respectively.
					\State  Update the dual variables ${{\boldsymbol{\mu}}_i}$ by (\ref{eq:eq51}).
				\Until the constraint violation indicator $\max \{\left \| \boldsymbol{\xi}-\mathbf{z}_1 \right \|_{F} ,\left \| \boldsymbol{\xi}-\mathbf{z}_2 \right \|_{F}  \}$ is smaller than $\epsilon _1$ or $t$ reaches the maximum number of iterations $T_1$.
				\EndIf 
		\State  Set $t=t+1$.
		\Until{ the objective function value, i.e., Eq. (\ref{eq:eq33a}) converges.}
	\end{algorithmic} 
	\end{algorithm}

\section{Numerical results}\label{se:se5}
In this section, we provide simulation results to demonstrate the effectiveness of the proposed sensing algorithm as well as the beamforming design algorithm in high-mobility systems. {Without loss of generality, we set the number of reflecting elements as $N_I=8\times8$ and the number of BS antennas as $N_B=4$.} The number of the paths from the user to the IRS and from the IRS to the user is set as $L_{\mathrm{UI}}=L_{\mathrm{IB}}=4$, and the carrier frequency is 28 $\mathrm{GHz}$ \cite{IRS-OTFS2}. {For} the OTFS modulation, the number of the subcarriers and time slots are $M=64$ and $N=16$, respectively. The subcarrier spacing is $\Delta f=15 \mathrm{kHz}$. Moreover, the normalized maximum delay tap is $l_{\mathrm{max}}=8$ and the delay taps are randomly generated within $[0,l_{\mathrm{max}}]$, while the maximum velocity of the user is $v_{\mathrm{max}}=120 \mathrm{km/h}$ with Doppler shifts generated by $\nu_i=v_{\mathrm{max}}\cos(\theta_i)$, where $\theta_i$ is randomly and uniformly generated within $[0,2\pi ]$ \cite{basic}.
Unless otherwise specified, we set $T_1=10$, $\epsilon _1=10^{-6}$, $\gamma_1=10^{-3}$, and $\rm{SNR} = \frac{{x_p^2}}{{{\sigma ^2}}}=20$ dB.
Simulation results are obtained by averaging over $1000$ independent trials.
\begin{figure}[htbp]
\centering
\includegraphics[width=3.2in]{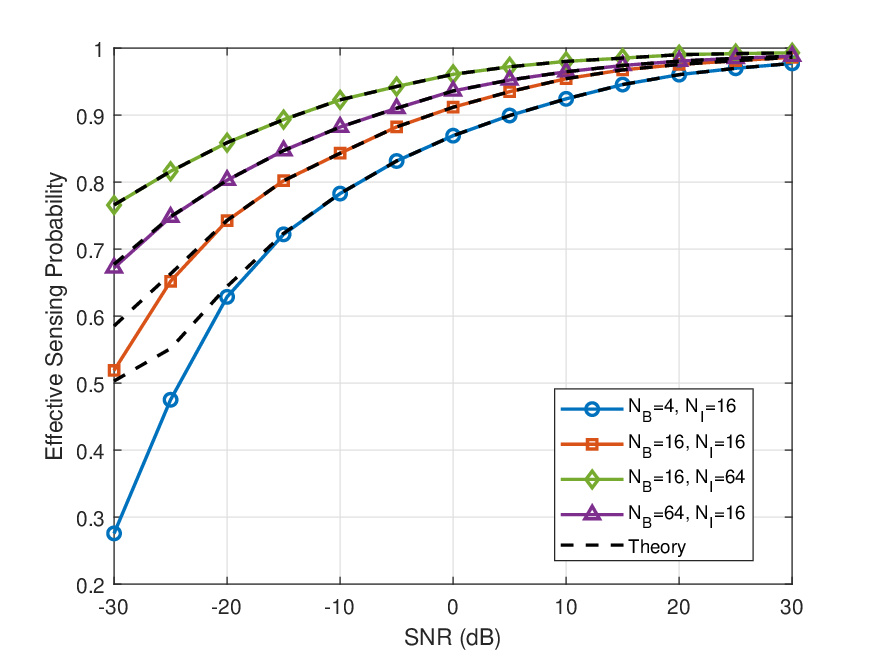}
\caption{Effective sensing probability of the proposed ratio-based sensing algorithm.}\label{fig:fig4}
\end{figure}
\subsection{{The Performance of the Proposed Sensing Method}}
Fig. \ref{fig:fig4} illustrates the asymptotic effective sensing probability derived in Eq. (\ref{eq:eq22}) from Theorem \ref{theorem:2}, across various numbers of BS antennas $N_{{B}}$ and reflecting elements $N_I$. The results highlight the robust performance of the proposed ratio-based sensing algorithm under different antenna setups.
Notably, the effective sensing probability closely matches the theoretical sensing probability, i.e., Eq. (\ref{eq:eqtp}),  with increasing SNR, $N_I$, and $N_B$, and eventually approaches 1. This underscores the reliability of algorithm: we can always guarantee successful sensing by increasing $N_I$, $N_B$ and SNR.   For instance, at an SNR of $10 \mathrm{dB}$, with $N_I=64$ and $N_B=16$, the effective sensing probability remains approximately 98.03\%.
\begin{figure}[h]
  \centering
  \subfigure[The impact of $N$ on the achievable MSE.]{\label{fig:fig5a}\includegraphics[width=3.2in]{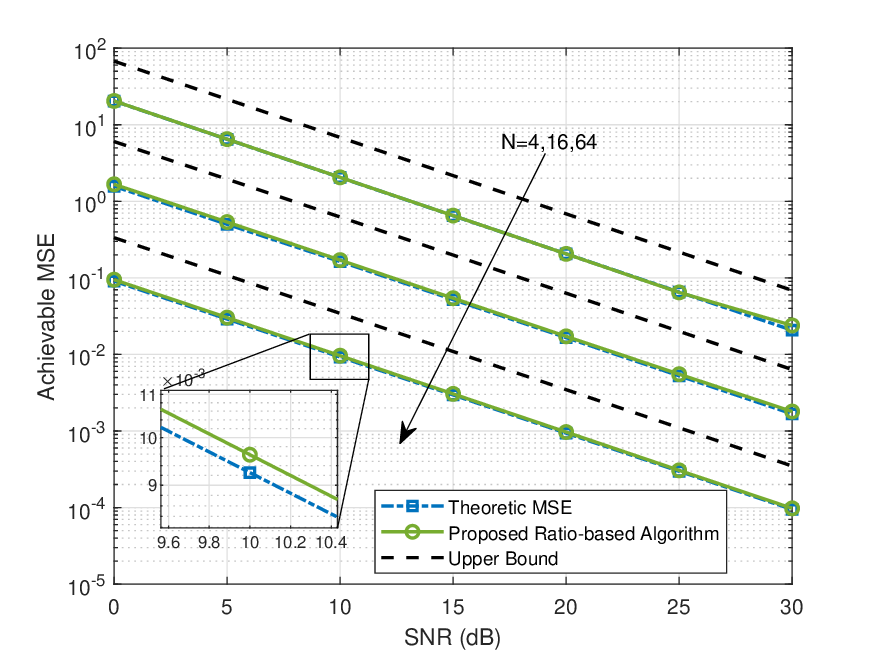}}
 \subfigure[The impact of $N_I$ on the achievable MSE.]{\label{fig:fig5b}\includegraphics[width=3.2in]{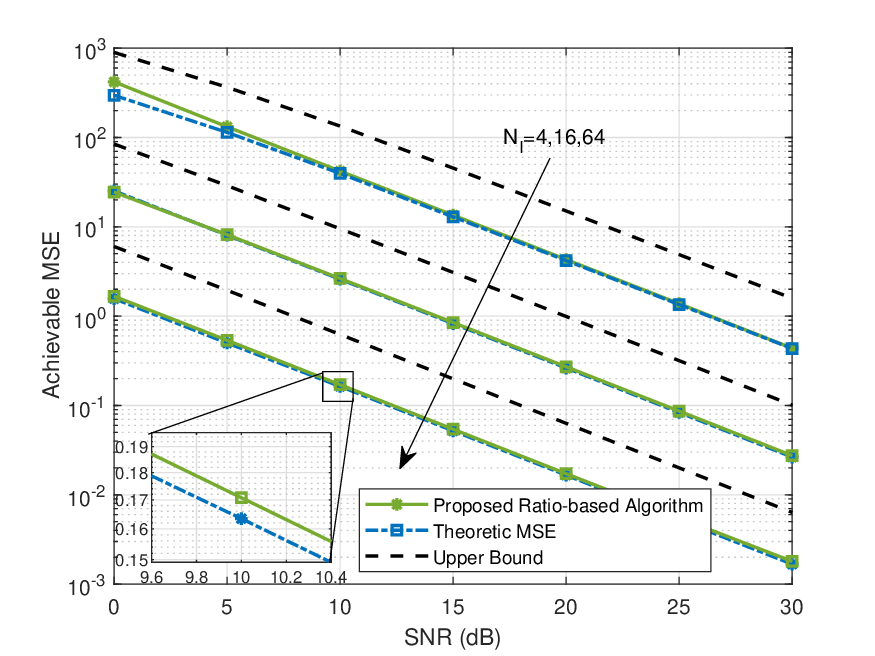}}
\caption{The achievable MSE of the proposed ratio-based sensing algorithm.}\label{fig:fig5}
\end{figure}

\begin{figure}[h]
  \centering
  \subfigure[Comparison of different sensing algorithms.]{\label{fig:fig6a}\includegraphics[width=3.2in]{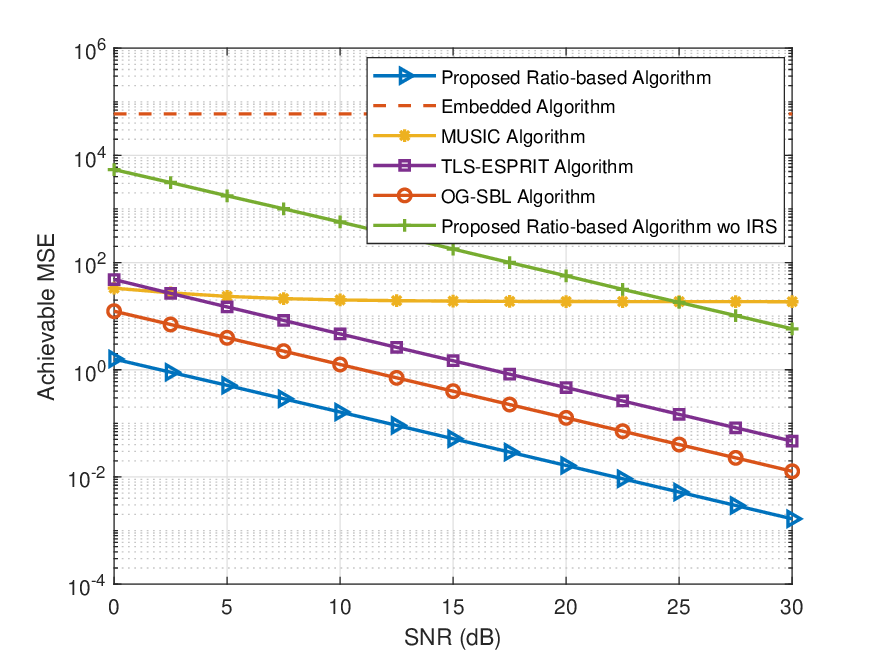}}
 \subfigure[The impact of the velocity on the sensing algorithms.]{\label{fig:fig6b}\includegraphics[width=3.2in]{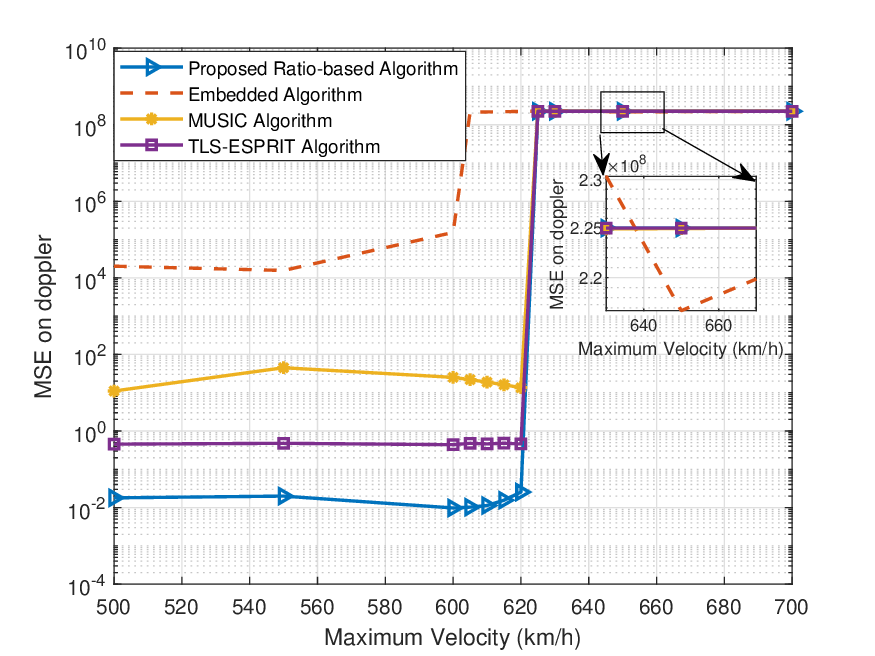}}
\caption{The achievable MSE of the proposed ratio-based sensing algorithm.}\label{fig:fig6}
\end{figure}

Fig. \ref{fig:fig5} evaluates the performance of the proposed ratio-based sensing algorithm, {where the ``Theoretic MSE'' and the ``Upper Bound'' curves are plotted by ${\varepsilon }    = \left[ {| \nu _1^{{\rm{UI}}} - \hat \nu _1^{{\rm{UI}}}{|^2}} \right]$ and Eq. (\ref{eq:eqeup}) in Corollary \ref{corollary:1}, respectively.} The results indicate that the approximate achievable MSE of the proposed algorithm tightly aligns with the theoretical MSE. Consistent with Remark \ref{re:remark:3}, the achievable MSE decreases with the increased resolution in the Doppler domain  (i.e., $\frac{1}{NT_s}$) and $N_I$. This highlights the significant role of the IRS in enhancing the sensing performance.

Fig. \ref{fig:fig6} compares the proposed sensing algorithm with various sensing algorithms, including the ``Embedded Algorithm'' \cite{embedded1}, the ``MUSIC Algorithm'', the ``OG-SBL Algorithm''  \cite{SBL_MUSIC},  and the ``TLS-ESPRIT Algorithm'' \cite{TLS-ESPRIT}. For the OG-SBL and MUSIC algorithms, the number of grids is 160 and $K_N$=1000, respectively. Specific complexity comparison for the proposed algorithm and benchmarks is shown in TABLE \ref{tab1}. 
The results indicate that the proposed algorithm has lower complexity as well as higher performance compared to these benchmarks. Notably, unlike the Embedded and MUSIC algorithms, which encounter a performance floor limited by the number of samples in the Doppler domain, the proposed algorithm is not subject to such constraints. 
This distinction underscores the superior efficiency and effectiveness of the proposed algorithm. 
Besides, our algorithm is designed for operation within a velocity range up to $\frac{\lambda}{T_s}$ m/s, which aligns with the typical velocity constraints in current OTFS-based systems\footnote{For OTFS systems,  when the maximum Doppler shift $\nu_1^{{\rm{UI}}}>\frac{1}{T_s}$, the corresponding Doppler index $k$ will exceed $N$, leading to estimation failures. Therefore, the maximum velocity of the user is  $\frac{\lambda}{T_s}$ m/s.}. Fig. \ref{fig:fig6b} shows the sensing performance at different velocities when $\mathrm{SNR}=20$ dB, where the maximum velocity of the user is about $v_{\mathrm{max}}$=623.0769 km/h. It is observed that all algorithms fail when the velocity exceeds $v_{\mathrm{max}}$. However, the maximum velocity surpasses the typical speeds of vehicles. Therefore, the proposed algorithm demonstrates broad applicability for high-mobility scenarios. Moreover, when up to this critical velocity, our algorithm demonstrates the highest estimation accuracy among the compared methods.
Therefore, the superiority of the proposed scheme lies in its low computational complexity and the ability to achieve high-precision sensing performance in scenarios of high-mobility with a single pilot symbol.
Moreover, we further illustrate the enhancement by IRS. In scenarios without IRS, the IRS is treated as a scatterer with a phase shift of $\pi$. Simulation demonstrates that the IRS can markedly enhance the sensing performance.

\begin{figure*}[htbp]
\centering
\begin{minipage}{0.32\linewidth}
\centering
\includegraphics[width=2.3in]{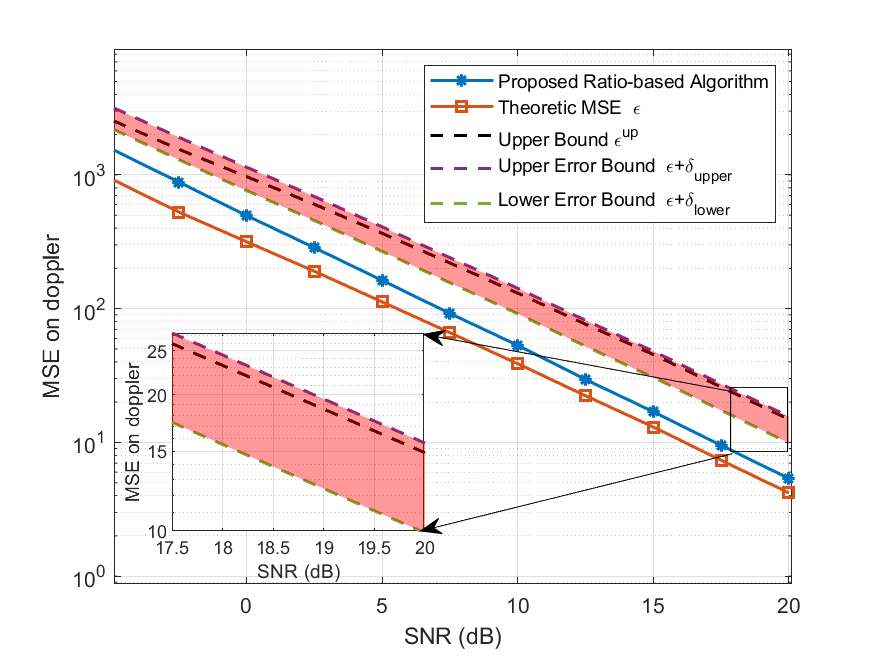}
\caption{The tightness evaluation of the upper bound of the achievable MSE.}\label{fig:figeb}
\end{minipage}
\begin{minipage}{0.32\linewidth}
\centering
\includegraphics[width=2.3in]{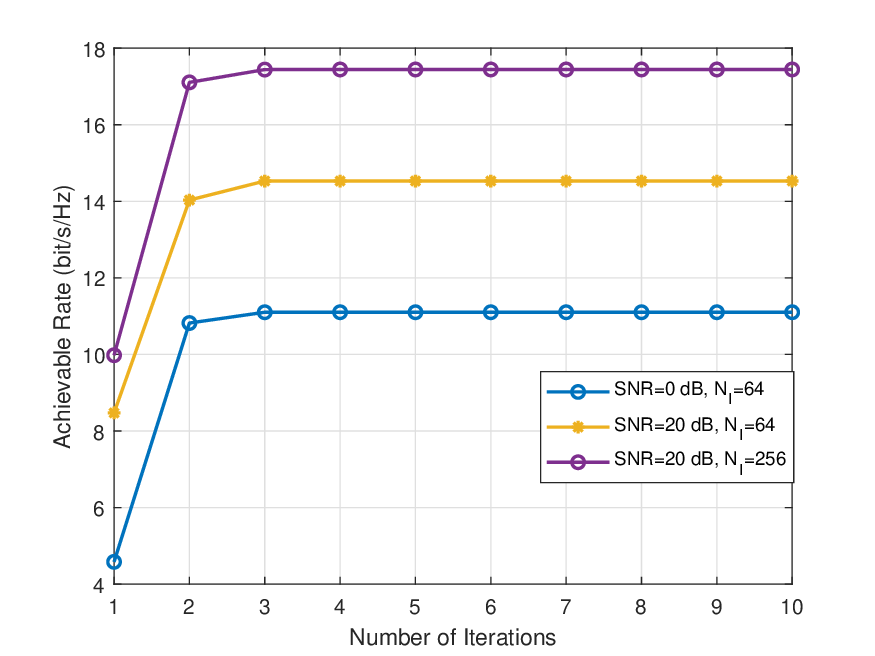}
\caption{The convergence of the proposed subspace-based algorithm.}\label{fig:fig7}
\end{minipage}
\begin{minipage}{0.32\linewidth}
\centering
\includegraphics[width=2.3in]{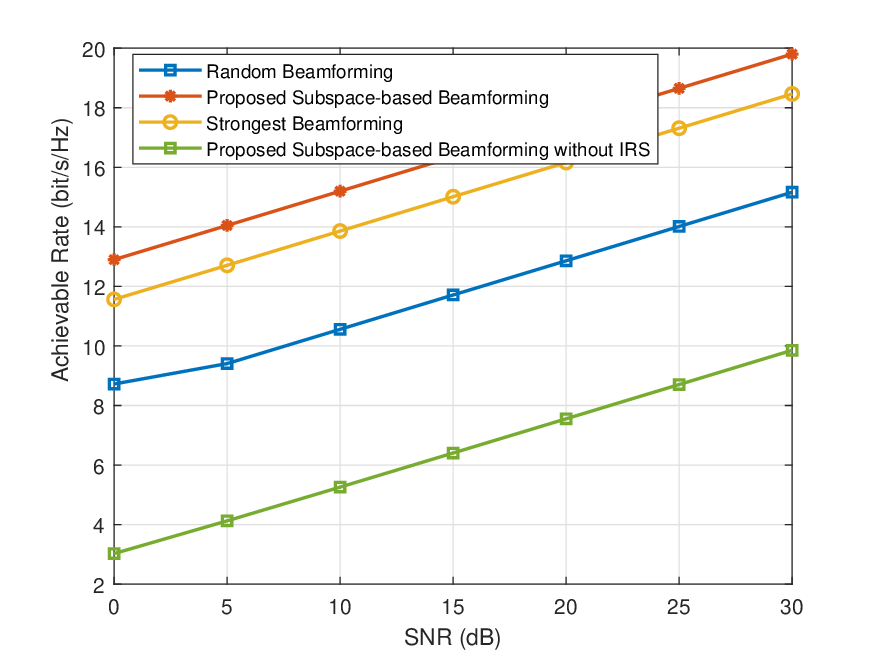}
\caption{Comparison of different beamforming design algorithms.}\label{fig:fig8}
\end{minipage}
\end{figure*}

Fig. \ref{fig:figeb} illustrate the tightness of the upper bound  for $N_B=16$ and $N_I=4$, where the ``Upper Error Bound'' and ``Lower Error Bound'' are defined as: $\varepsilon +\delta_{\rm{upper}}$ and $\varepsilon +\delta_{\rm{lower}}$, respectively.
It can be observed from the figure that the discrepancy between the derived achievable MSE and its upper bound aligns closely with the expected error margins, i.e., $\varepsilon+\delta_{\rm{lower}} < {\varepsilon ^{{\rm{up}}}}  <\varepsilon+\delta_{\rm{upper}}$, emphasizing the precision of the established error bounds.

%

\subsection{{The Performance of the Proposed Beamforming Design Algorithm}}
So far we have been discussing the sensing performance. In this subsection, we show the performance of the proposed beamforming design algorithm.

We first present the convergence properties of the proposed subspace-based beamforming algorithm, as depicted in Fig \ref{fig:fig7}. It can be observed that the proposed beamforming design algorithm can achieve convergence in a small iteration number, typically within three iterations. This can be attributed to the closed-form solution, which significantly expedites the convergence process.

Fig. \ref{fig:fig8} compares the communication performance across various beamforming algorithms. In the ``Random Beamforming'' algorithm,  IRS phase shifts are randomly and uniformly selected from the range $[0,2\pi]$, combined with our proposed algorithm at the BS. The ``Strongest Beamforming'' scheme proposed in [34], serves as our benchmark. It optimizes the IRS phase shift matrix to maximize the signal power of the LoS cascaded paths, while overlooking interference from other paths, i.e., maximize $\left| {h_{1,1}^{{\rm{UIB}}}}\right|$. And the combining vector is designed using our proposed subspace-based algorithm. The results clearly show that the proposed subspace-based algorithm outperforms these baseline algorithms, owing to the utilization of the path diversity among the BS, the IRS, and the user.

\section{Conclusion}\label{se:se6}
This paper introduced an IRS-assisted ISAC framework for high-mobility systems, utilizing OTFS modulation to harness the DD spread. The proposed framework incorporates the ratio-based velocity estimation algorithm and a subspace-based beamforming design algorithm. We evaluated the performance of the proposed sensing algorithm from the perspective of effective sensing probability and achievable MSE. The estimation accuracy depends on the number of reflecting elements and the Doppler domain resolution. Furthermore, the proposed beamforming design algorithm significantly enhances the communication capabilities while ensuring robust sensing performance. Simulation results demonstrate that the proposed subspace-based beamforming design algorithm outperforms the existing state-of-the-art ones that design the IRS phase shifts aligning the strongest path.
Future research could productively extend the proposed algorithms to varied settings, such as multi-cell or cell-free IRS-assisted ISAC systems with high mobility, and analyze the sensing as well as the communication performance. Another promising research area involves the development of high-precision and low-complexity sensing algorithms that accommodate practical scenarios such as the system with finite delay resolution, where fractional delays will impact the performance of the system.

\appendices
\section{Proof of Theorem \ref{theorem:1}}\label{app:theorem:1}
Recall Eq. (\ref{eq:eq18}), when $k=\nu_1^{{\rm{UI}}}NT_s+k_p$, the maximum value of $\mathbf{Z}_{k,l}$ can be obtained, however, we can not always guarantee that there exists an integer Doppler index $k$ to satisfy this condition due to the fractional Doppler shift. However, we can guarantee that $k_1$ can always make ${\hat \nu_1}$ satisfy ${\hat \nu_1}\in\left[ -\frac{1}{2NT_s},0\right)$, if $\mathbf{Z}_{k_2,l}'>\mathbf{Z}_{k_3,l}'$ and ${\hat \nu_1}\in\left( 0,\frac{1}{2NT_s}\right]$, if $\mathbf{Z}_{k_2,l}'<\mathbf{Z}_{k_3,l}'$. 

We first consider the condition when $\mathbf{Z}_{k_2,l}'>\mathbf{Z}_{k_3,l}'$, and it can be observed that 
\begin{equation}\small
\begin{array}{l}
\sin \left( { - \frac{\pi }{{2N}}} \right) \le \sin \left( {\pi T_s{{\hat v}_1}} \right) <
0 <  \sin \left( {\pi T\left( {{{\hat v}_1} + \frac{1}{{NT_s}}} \right)} \right) < \sin \left( {\frac{\pi }{N}} \right).
\end{array}
\end{equation}

Therefore, we have the ratio of $\mathbf{Z}_{k_1,l}'$ and $\mathbf{Z}_{k_2,l}'$ as
\begin{equation}
\frac{\mathbf{Z}_{k_2,l}'}{\mathbf{Z}_{k_1,l}'}=\left| {\frac{{\sin \left( {\pi T_s{{\hat \nu }_1}} \right)}}{{\sin \left( {\pi T_s\left( {{{\hat \nu }_1} + \frac{1}{{NT_s}}} \right)} \right)}}} \right| \overset{(a)}= -{\frac{{\sin \left( {\pi T_s{{\hat \nu }_1}} \right)}}{{\sin \left( {\pi T_s\left( {{{\hat \nu }_1} + \frac{1}{{NT_s}}} \right)} \right)}}},
\end{equation}
where (a) exploit the fact that the two points satisfy $-\frac{1}{2NT_s}\le {\hat \nu_1} <0<{\hat \nu_2}={\hat \nu_1} + \frac{1}{{NT_s}}$. After some mathematical transformations, the above equation can be simplified as
\begin{equation}
\sin \left( {\pi T_s{{\hat v}_1}} \right)\cos \psi  + \cos \left( {\pi T_s{{\hat v}_1}} \right)\sin \psi  = \sin \left( {\pi T_s{{\hat v}_1} + \psi } \right) = 0,
\end{equation}
where $\tan \psi  = \frac{{\sin \left( {\frac{\pi }{N}} \right)\mathbf{Z}_{k_2,l}'}}{{\mathbf{Z}_{k_1,l}' + \mathbf{Z}_{k_2,l}'\cos \left( {\frac{\pi }{N}} \right)}}$.

Similarly, for the condition when $\mathbf{Z}_{k_2,l}'<\mathbf{Z}_{k_3,l}'$, it can be proved that
\begin{equation}\small
\begin{array}{l}
\hspace{-2mm}\sin \left( {\pi T_s\left( {{{\hat v}_1} - \frac{1}{{NT_s}}} \right)} \right) < \sin \left( { - \frac{\pi }{{2N}}} \right) < 0  \le \sin \left( {\pi T_s{{\hat v}_1}} \right) < \sin \left( {\frac{\pi }{{2N}}} \right),
\end{array}
\end{equation}
and the ratio of $\mathbf{Z}_{k_1,l}'$ and $\mathbf{Z}_{k_3,l}'$ is given by
\begin{equation}
\frac{\mathbf{Z}_{k_3,l}'}{\mathbf{Z}_{k_1,l}'} = \left| {\frac{{\sin \left( {\pi T_s{{\hat v}_1}} \right)}}{{\sin \left( {\pi T_s\left( {{{\hat v}_1} - \frac{1}{{NT_s}}} \right)} \right)}}} \right|=  - \frac{{\sin \left( {\pi T{{\hat v}_1}} \right)}}{{\sin \left( {\pi T\left( {{{\hat v}_1} - \frac{1}{{NT}}} \right)} \right)}},
\end{equation}
which can be simplified as 
\begin{equation}
\sin \left( {\pi T_s{{\hat v}_1}} \right)\cos \psi  + \cos \left( {\pi T_s{{\hat v}_1}} \right)\sin \psi  = \sin \left( {\pi T_s{{\hat v}_1} + \psi } \right) = 0,
\end{equation}
where $\tan \psi  = -\frac{{\sin \left( {\frac{\pi }{N}} \right)\mathbf{Z}_{k_3,l}'}}{{\mathbf{Z}_{k_1,l}' + \mathbf{Z}_{k_3,l}'\cos \left( {\frac{\pi }{N}} \right)}}$, and this yields Eq. (\ref{eq:eq19}).

\section{Proof of Theorem \ref{theorem:2}}\label{app:theorem:2}
Without loss of generality, we focus on the condition that $\mathbf{Z}_{k_2,l}'>\mathbf{Z}_{k_3,l}'$ as an example. In the high SNR regime, the power is mainly concentrated on the main lobe, thus $\mathrm{Pr}\left[\left(\mathbf{Z}_{k_1,l}\geq\max_{k}\mathbf{Z}_{k,l}\right)\right] \approx 1$ is always satisfied. And the effective sensing probability can be rewritten as
\begin{equation}
\begin{aligned}
\mathrm{P}_{\mathrm{eff}}\approx \mathrm{P}_{\mathrm{eff}}^{c_1}= \mathrm{Pr}\left[\frac{\mathbf{Z}_{k_2,l}}{\mathbf{Z}_{k_3,l}}> 1\right]. 
\end{aligned}
\end{equation}

Recall $\mathbf{Z}_{k,l}\triangleq \left | {\bf{Y}}^{\mathrm{DD}}_{\rm{p'}}\left[ {k,{l_p} + {l_{{\tau _1}}^{\mathrm{UI}}} + {l_{{\tau _1}}^{\mathrm{IB}}}} \right] \right |  $, and it follows Rician distribution with complicated probability density function (PDF), making the analysis untrackable. However, $\mathbf{Z}_{k,l}^2$ follows the non-central chi-square distribution with the degree of two, which can be approximated as a gamma distribution \cite{gamma}. Therefore, $\mathbf{Z}_{k,l}$ can be approximated as a Nakagami distribution \cite{nakagami}, i.e., $\mathbf{Z}_{k,l}\sim \mathrm{Na}({\varpi ,\Omega })$ with the following PDF:
\begin{equation}
\begin{aligned}
f(x;\varpi ,\Omega )& = \frac{{2{\varpi ^\varpi }}}{{\Gamma (\varpi ){\Omega ^\varpi }}}{x^{2\varpi  - 1}}\exp ( - \frac{\varpi }{\Omega }{x^2}),\forall x \ge 0.\\
& = \frac{{2{\vartheta ^\varpi }}}{{\Gamma (\varpi )}}{x^{2\varpi  - 1}}\exp ( - \vartheta {x^2}),\forall x \ge 0,
\end{aligned}
\end{equation} 
where $\varpi  = \frac{{{{ {{\rm{E}}^2\left[ {{\left(\mathbf{Z}_{k,l}'\right)^2}} \right]} }}}}{{{\rm{Var}}\left[ {{\left(\mathbf{Z}_{k,l}'\right)^2}} \right]}} = \frac{{{{\left[ {{\left(\mathbf{Z}_{k,l}'\right)^2} + {\sigma ^2}} \right]}^2}}}{{2{\left(\mathbf{Z}_{k,l}'\right)^2}{\sigma ^2} + {\sigma ^4}}}$, $\Omega  = {\rm{E}}\left[ {{\left(\mathbf{Z}_{k,l}'\right)^2}} \right] = {\left(\mathbf{Z}_{k,l}'\right)^2} + {\sigma ^2}$, and $\vartheta  = \frac{\varpi }{\Omega }$.

Define $f_{\rm{R}_{{{\rm{}}_1},{{\rm{}}_2}}} = \frac{\mathbf{Z}_{k_1,l}'}{\mathbf{Z}_{k_2,l}'}$, the PDF can be derived according to \cite[Eq. 3.326.2]{pdf}, as
\begin{equation}\label{eq:eq34}
\begin{aligned}
{f_{\rm{R}_{{{\rm{}}_1},{{\rm{}}_2}}}}(r) &= \int_0^\infty  {\left| t \right|} {f_{\mathbf{Z}_{k_1,l}'}}(rt){f_{\mathbf{Z}_{k_2,l}'}}(t){\rm{d}}t\\
&  = \frac{{2\vartheta _2^{{\varpi _2}}\vartheta _1^{{\varpi _1}}\Gamma ({\varpi _2} + {\varpi _1})}}{{\Gamma ({\varpi _2})\Gamma ({\varpi _1})}}\frac{{{r^{2{\varpi _1} - 1}}}}{{{{\left( {{\vartheta _1}{r^2} + {\vartheta _2}} \right)}^{{\varpi _2} + {\varpi _1}}}}}.
\end{aligned}
\end{equation}

Therefore, $\mathrm{P}_{\mathrm{eff}}^{c_1}$ can be regarded as the cumulative distribution function (CDF) of $f_{R_{{{\rm{}}_1},{{\rm{}}_2}}}$ and can be derived according to \cite[Eq. 3.194.2]{pdf}, yields Eq. (\ref{eq:eq22}). 

\section{Proof of Corollary \ref{corollary:1}}\label{app:corollary:1}
Without loss of generality, we focus on the condition that $\mathbf{Z}_{k_2,l}'>\mathbf{Z}_{k_3,l}'$ as an example. Recall Eq. (\ref{eq:eq19}) $\sim$ (\ref{eq:eq22}), the achievable MSE can be rewritten as
\begin{equation}
\begin{aligned}
\varepsilon & = \frac{1}{{{\pi ^2}{T_s^2}}}\left[ {|\psi  - \psi '{|^2}} \right]
 \overset{(a)} \approx \frac{{{{\sin }^4}\psi '}}{{{\pi ^2}{T^2}}}\left[ {|\cot \psi  - \cot \psi '{|^2}} \right],
\end{aligned}
\end{equation}
where (a) exploits the first-order approximation $\frac{{\cot \psi  - \cot \psi '}}{{\psi  - \psi '}} \approx  - \frac{1}{{{{\sin }^2}\psi '}},\psi ' \ne 0,$ to remove the arccot function\footnote{The accuracy of the approximation increases with the decreasing of the value $|\psi-\psi'|$}. Moreover, taking $\cot \psi  = \frac{{{\mathbf{Z}_{k_1,l}}}}{{\sin \left( {\frac{\pi }{N}} \right){\mathbf{Z}_{k_2,l}}}} + \cot \left( {\frac{\pi }{N}} \right)$ into the above equation, $\varepsilon$ and $\cot \psi ' = \frac{{{\mathbf{Z}'_{k_1,l}}}}{{\sin \left( {\frac{\pi }{N}} \right){\mathbf{Z}'_{k_2,l}}}} + \cot \left( {\frac{\pi }{N}} \right)$ into the above equation, $\varepsilon$ can be written as
\begin{equation}
\varepsilon = \frac{{{{\sin }^4}\psi '}}{{{{\sin }^2}\left( {\frac{\pi }{N}} \right){\pi ^2}{T_s^2}}}\left\{ {{{ {\frac{{{\mathbf{Z}'^2_{k_1,l}}}}{{{\mathbf{Z}'^2_{k_2,l}}}}}}} - 2 {\frac{{{\mathbf{Z}'_{k_1,l}}}}{{{\mathbf{Z}'_{k_2,l}}}}} \mathbb{E}\left( {\frac{{{\mathbf{Z}_{k_1,l}}}}{{{\mathbf{Z}_{k_2,l}}}}} \right) + \mathbb{E}{{\left( {\frac{{{\mathbf{Z}^2_{k_1,l}}}}{{{\mathbf{Z}^2_{k_2,l}}}}} \right)}}} \right\},
\end{equation}
where $\mathbb{E}(x)$ denotes the expectation of the variable $x$. According to \cite[Eq. 3.194.1]{pdf}, the expectation can be further calculated as
\begin{equation}
\begin{aligned}
{\frac{{\mathbf{Z}_{k_1,l}}}{{\mathbf{Z}_{k_2,l}}}} &= \int_0^\infty  {{f_{{\rm{R}}{{\rm{}}_{{I_1},{I_2}}}}}} (r)r{\rm{d}}r
 &= \frac{{\Gamma ({\varpi _2} - \frac{1}{2})\Gamma ({\varpi _1} + \frac{1}{2})}}{{\Gamma ({\varpi _2})\Gamma ({\varpi _1})}}\sqrt {\frac{{{\vartheta _2}}}{{{\vartheta _1}}}}, 
\end{aligned}
\end{equation} 
and 
\begin{equation}
\begin{aligned}
 {\frac{{\mathbf{Z}^2_{k_1,l}}}{{\mathbf{Z}^2_{k_2,l}}}} &= \int_0^\infty  {{f_{{\rm{R}}{{\rm{}}_{{I_1},{I_2}}}}}} (r){r^2}{\rm{d}}r
 &= \frac{{{\varpi _1}}}{{{\varpi _2} - 1}}\frac{{{\vartheta _2}}}{{{\vartheta _1}}},
\end{aligned}
\end{equation}
this yields Eq. (\ref{eq:eq25}). 
The work \cite{errorbound} derived a more compact upper bound and gave the ratio bound as
\begin{equation}
1 < \frac{{\Gamma \left( x \right)}}{{\sqrt {2\pi } {x^{x - \frac{1}{2}}}{e^{ - x}}}} < 1 + \frac{1}{{12x}} + \frac{1}{{288{x^2}}},
\end{equation}
based on which, the lower bound of the ratio function can be derived as Eq. (\ref{eq:eq73}),
\begin{figure*}
\begin{equation}\label{eq:eq73}
\small
\begin{aligned}
\frac{{\Gamma \left( {{\varpi _2} - \frac{1}{2}} \right)\Gamma \left( {{\varpi _1} + \frac{1}{2}} \right)}}{{\Gamma \left( {{\varpi _2}} \right)\Gamma \left( {{\varpi _1}} \right)}} &> \frac{{{{\left( {{\varpi _2} - \frac{1}{2}} \right)}^{ - (1/2) + \left( {{\varpi _2} - \frac{1}{2}} \right)}}{e^{ - \left( {{\varpi _2} - \frac{1}{2}} \right)}}{{(2\pi )}^{1/2}}{{\left( {{\varpi _1} + \frac{1}{2}} \right)}^{ - (1/2) + \left( {{\varpi _1} + \frac{1}{2}} \right)}}{e^{ - \left( {{\varpi _1} + \frac{1}{2}} \right)}}{{(2\pi )}^{1/2}}}}{{{{(2\pi )}^{1/2}}\varpi _2^{ - (1/2) + {\varpi _2}}{e^{ - {\varpi _2}}}{{(2\pi )}^{1/2}}\varpi _1^{ - (1/2) + {\varpi _1}}{e^{ - {\varpi _1}}}\left( {1 + \frac{1}{{12{\varpi _1}}} + \frac{1}{{288{\varpi _1}}}} \right)\left( {1 + \frac{1}{{12{\varpi _2}}} + \frac{1}{{288{\varpi _2}}}} \right)}}\\
&{\rm{ = }}{\left( {1 - \frac{1}{{2{\varpi _2}}}} \right)^{ - 1 + {\varpi _2}}}{\left( {1 + \frac{1}{{2{\varpi _1}}}} \right)^{{\varpi _1}}}\sqrt {\frac{{{\varpi _1}}}{{{\varpi _2}}}} \frac{1}{{\left( {1 + \frac{1}{{12{\varpi _1}}} + \frac{1}{{288{\varpi _1}}}} \right)\left( {1 + \frac{1}{{12{\varpi _2}}} + \frac{1}{{288{\varpi _2}}}} \right)}}\\
&\mathop  \approx \limits^{(a)} \sqrt {\frac{{{\varpi _1}}}{{{\varpi _2}}}},
\end{aligned}
\end{equation}\hrulefill
\end{figure*}
where (a) holds due to the large value of ${\varpi _1}$ and ${\varpi _2}$. Then, the upper bound of achievable MSE can be derived.


\section{Tightness analysis of the upper bound (\ref{eq:eq26})}\label{app:Tightness}
Firstly, the ratio between  ${\frac{{\Gamma \left( {{\varpi _2} - \frac{1}{2}} \right)\Gamma \left( {{\varpi _1} + \frac{1}{2}} \right)}}{{\Gamma \left( {{\varpi _2}} \right)\Gamma \left( {{\varpi _1}} \right)}}}$ and its approximate value ${\sqrt {\frac{{{\varpi _1}}}{{{\varpi _2}}}} }$  can be reformed as
\begin{equation}\small
\frac{{\frac{{\Gamma \left( {{\varpi _1} + \frac{1}{2}} \right)}}{{\Gamma \left( {{\varpi _1}} \right)}}\frac{{\Gamma \left( {{\varpi _2} - \frac{1}{2}} \right)}}{{\Gamma \left( {{\varpi _2}} \right)}}}}{{\sqrt {\frac{{{\varpi _1}}}{{{\varpi _2}}}} }}={\kappa _1}\left( {{\varpi _1}} \right) \cdot {\kappa _2}\left( {{\varpi _2}} \right) \cdot {\kappa _3}\left( {{\varpi _1},{\varpi _2}} \right),
\end{equation}
where 

${\kappa _1}\left( {{\varpi _1}} \right){\rm{ = }}{\mu _1}\left( {{\varpi _1}} \right)\frac{{\Gamma \left( {{\varpi _1} + \frac{1}{2}} \right)}}{{\Gamma \left( {{\varpi _1}} \right)\sqrt {{\varpi _1}} }}, {\kappa _2}\left( {{\varpi _2}} \right){\rm{ = }}{\mu _2}\left( {{\varpi _2}} \right)\frac{{\Gamma \left( {{\varpi _2} - \frac{1}{2}} \right)}}{{\Gamma \left( {{\varpi _2}} \right)\sqrt {\frac{1}{{{\varpi _2}}}} }}$,
${\kappa _3}\left( {{\varpi _1},{\varpi _2}} \right){\rm{ = }}\frac{1}{{{\mu _1}\left( {{\varpi _1}} \right){\mu _2}\left( {{\varpi _2}} \right)}}$,
 ${\mu _1}\left( {{\varpi _1}} \right){\rm{ = }}\frac{{\left( {1 + \frac{1}{{12{\varpi _1}}} + \frac{1}{{288{\varpi _1}}}} \right)}}{{{{\left( {1 + \frac{1}{{2{\varpi _1}}}} \right)}^{{\varpi _1}}}}}$, and ${\mu _2}\left( {{\varpi _2}} \right) = \frac{{\left( {1 + \frac{1}{{12{\varpi _2}}} + \frac{1}{{288{\varpi _2}}}} \right)}}{{{{\left( {1 - \frac{1}{{2{\varpi _2}}}} \right)}^{ - 1 + {\varpi _2}}}}}$.

According to \cite{errorbound}, the ratio error bound between  ${\Gamma \left( x \right)}$ and its approximate value ${{\sqrt {2\pi } {x^{x - \frac{1}{2}}}{e^{ - x}}}}$ is given by
\begin{equation}\small
1 < \frac{{\Gamma \left( x \right)}}{{\sqrt {2\pi } {x^{x - \frac{1}{2}}}{e^{ - x}}}} < 1 + \frac{1}{{12x}} + \frac{1}{{288{x^2}}},
\end{equation}
and
\begin{equation}\small
1 < \frac{{\sqrt {2\pi } {x^{x - \frac{1}{2}}}{e^{ - x}}\left( {1 + \frac{1}{{12x}} + \frac{1}{{288{x^2}}}} \right)}}{{\Gamma \left( x \right)}} < 1 + \frac{1}{{12x}} + \frac{1}{{288{x^2}}}.
\end{equation}

Thus the ratio error bound between ${\frac{{\Gamma \left( {{\varpi _2} - \frac{1}{2}} \right)}}{{\Gamma \left( {{\varpi _2}} \right)}}}$ and its approximate value ${{\frac{{{{(2\pi )}^{1/2}}{{\left( {{\varpi _2} - \frac{1}{2}} \right)}^{ - (1/2) + \left( {{\varpi _2} - \frac{1}{2}} \right)}}{e^{ - \left( {{\varpi _2} - \frac{1}{2}} \right)}}}}{{{{(2\pi )}^{1/2}}\varpi _2^{ - (1/2) + {\varpi _2}}{e^{ - {\varpi _2}}}\left( {1 + \frac{1}{{12{\varpi _1}}} + \frac{1}{{288{\varpi _1}}}} \right)}}}} $ can be derived as
\begin{equation}\small
\begin{array}{l}
1 < \frac{{\frac{{\Gamma \left( {{\varpi _2} - \frac{1}{2}} \right)}}{{\Gamma \left( {{\varpi _2}} \right)}}}}{{\frac{{{{(2\pi )}^{1/2}}{{\left( {{\varpi _2} - \frac{1}{2}} \right)}^{ - (1/2) + \left( {{\varpi _2} - \frac{1}{2}} \right)}}{e^{ - \left( {{\varpi _2} - \frac{1}{2}} \right)}}}}{{{{(2\pi )}^{1/2}}\varpi _2^{ - (1/2) + {\varpi _2}}{e^{ - {\varpi _2}}}\left( {1 + \frac{1}{{12{\varpi _2}}} + \frac{1}{{288{\varpi _2}}}} \right)}}}} =\frac{{\kappa _2}\left( {{\varpi _2}} \right)}{e^{\frac{1}{2}}}\\
< \left( {1 + \frac{1}{{12\left( {{\varpi _2} - \frac{1}{2}} \right)}} + \frac{1}{{288\left( {{\varpi _2} - \frac{1}{2}} \right)}}} \right)\left( {1 + \frac{1}{{12{\varpi _2}}} + \frac{1}{{288{\varpi _2}}}} \right),
\end{array}
\end{equation}
similarly,   the ratio error bound between ${\frac{{\Gamma \left( {{\varpi _1} +\frac{1}{2}} \right)}}{{\Gamma \left( {{\varpi _1}} \right)}}}$ and its approximate value ${{\frac{{{{(2\pi )}^{1/2}}{{\left( {{\varpi _1} + \frac{1}{2}} \right)}^{ - (1/2) + \left( {{\varpi _1} + \frac{1}{2}} \right)}}{e^{ - \left( {{\varpi _1} + \frac{1}{2}} \right)}}}}{{{{(2\pi )}^{1/2}}\varpi _1^{ - (1/2) + {\varpi _1}}{e^{ - {\varpi _1}}}\left( {1 + \frac{1}{{12{\varpi _1}}} + \frac{1}{{288{\varpi _1}}}} \right)}}}}$ can be derived as
\begin{equation}\small
\begin{array}{l}
1 < \frac{{\frac{{\Gamma \left( {{\varpi _1} + \frac{1}{2}} \right)}}{{\Gamma \left( {{\varpi _1}} \right)}}}}{{\frac{{{{(2\pi )}^{1/2}}{{\left( {{\varpi _1} + \frac{1}{2}} \right)}^{ - (1/2) + \left( {{\varpi _1} + \frac{1}{2}} \right)}}{e^{ - \left( {{\varpi _1} + \frac{1}{2}} \right)}}}}{{{{(2\pi )}^{1/2}}\varpi _1^{ - (1/2) + {\varpi _1}}{e^{ - {\varpi _1}}}\left( {1 + \frac{1}{{12{\varpi _1}}} + \frac{1}{{288{\varpi _1}}}} \right)}}}} = {{\kappa _1}\left( {{\varpi _1}} \right)}{e^{\frac{1}{2}}}\\
< \left( {1 + \frac{1}{{12\left( {{\varpi _1} + \frac{1}{2}} \right)}} + \frac{1}{{288\left( {{\varpi _1} + \frac{1}{2}} \right)}}} \right)\left( {1 + \frac{1}{{12{\varpi _1}}} + \frac{1}{{288{\varpi _1}}}} \right).
\end{array}
\end{equation}
Then,  the ratio error bound between $\frac{{\Gamma \left( {{\varpi _2} - \frac{1}{2}} \right)\Gamma \left( {{\varpi _1} + \frac{1}{2}} \right)}}{{\Gamma \left( {{\varpi _2}} \right)\Gamma \left( {{\varpi _1}} \right)}}$ and $ \sqrt {\frac{{{\varpi _1}}}{{{\varpi _2}}}}$ can be derived as 
\begin{equation}\small
\alpha_{\rm{lower}} < \frac{{\frac{{\Gamma \left( {{\varpi _2} - \frac{1}{2}} \right)\Gamma \left( {{\varpi _1} + \frac{1}{2}} \right)}}{{\Gamma \left( {{\varpi _2}} \right)\Gamma \left( {{\varpi _1}} \right)}}}}{{\sqrt {\frac{{{\varpi _1}}}{{{\varpi _2}}}} }} < \alpha_{\rm{upper}}.
\end{equation}

Therefore, the error bound of the proposed upper bound of the achievable MSE can be derived.


\begin{thebibliography}{99}
\bibitem{ISAC1}
F. Liu, Y. Cui, C. Masouros, J. Xu, T. Han, Y. C. Eldar, and S. Buzzi, ``Integrated sensing and communications: Toward dual-functional wireless networks for 6G and beyond,'' {\em IEEE J. Sel. Areas Commun.}, vol. 40, no. 6, pp. 1728--1767, Jun. 2022.

\bibitem{ISAC2}
A. Zhang, M. L. Rahman, X. Huang, Y. J. Guo, S. Chen, and R. W. Heath, ``Perceptive mobile network: Cellular networks with radio vision via joint communication and radar sensing,'' {\em IEEE Veh. Technol. Mag.}, vol. 16, no. 2, pp. 20--30, Jun. 2021.

\bibitem{ISAC3}
J. A. Zhang, M. L. Rahman, K. Wu, X. Huang, Y. J. Guo, S. Chen, and J. Yuan, ``Enabling joint communication and radar sensing in mobile networks survey,'' {\em Commun. Survs \& Tuts}, vol. 24, no. 1, pp. 306--345, 2022.

\bibitem{ISAC4}
Q. Zhang, H. Sun, X. Gao, X. Wang, and Z. Feng, ``Time-division ISAC enabled connected automated vehicles cooperation algorithm design and performance evaluation,'' {\em IEEE J. Sel. Areas Commun.}, vol. 40, no. 7, pp. 2206--2218, Jul. 2022.




\bibitem{embedded1}
P. Raviteja, K. T. Phan, and Y. Hong,  ``Embedded pilot-aided channel estimation for OTFS
in delay–doppler channels,''  {\em IEEE Trans. Veh. Technol.}, vol. 68, no. 5, pp. 4906--4917, May 2019.

\bibitem{potential}
X. Xia, K. Xu, Y. Wang, Y. Xu, and W. Xie, ``Achieving better accuracy with less computations: A Delay-Doppler spectrum matching assisted active sensing framework for OTFS based ISAC systems,'' {\em IEEE Trans. Wireless Commun.}, early access, doi: 10.1109/TWC.2023.3330845.



\bibitem{IRS}
Q. Wu and R. Zhang, ``Towards smart and reconfigurable environment: Intelligent reflecting surface aided wireless network,'' {\em IEEE Commun. Mag.}, vol. 58, no. 1, pp. 106--112, Jan. 2020.


\bibitem{IRS-ISAC1}
Z. Xing, R. Wang, and X. Yuan, ``Joint active and passive beamforming design for reconfigurable intelligent surface enabled integrated sensing and communication,'' {\em IEEE Trans. Commun.}, vol. 71, no. 4, pp. 2457--2474, Apr. 2023.

\bibitem{IRS-ISAC2}
M. Hua, Q. Wu, C. He, S. Ma, and W. Chen, ``Joint active and passive beamforming design for IRS-aided radar-communication,'' {\em IEEE Trans. Wireless Commun.}, vol. 22, no. 4, pp. 2278--2294, Apr. 2023.

\bibitem{IRS-ISAC6}
C. Liao, F. Wang, and V. K. N. Lau, ``Optimized design for IRS-assisted integrated sensing and communication systems in clutter environments,'' {\em IEEE Trans. Commun.}, vol. 71, no. 8, pp. 4721--4734, Aug. 2023.

\bibitem{IRS-IASC-vehicular1}
H. Zhang, R. Liu, M. Li, W. Wang, and Q. Liu, ``Joint sensing and communication optimization in target-mounted STARS-assisted vehicular networks: A MADRL approach,’’ {\em IEEE Trans. Veh. Technol.}, early access, Feb. 2024, doi: 10.1109/TVT.2024.3364761.

\bibitem{IRS-IASC-vehicular2}
M. Li, S. Zhang, Y. Ge, Z. Li, F. Gao, and P. Fan, ``Integrated sensing and communication with STAR-RIS over high mobility scenario,’’ in {\em Proc. IEEE Global Commun. Conf. (GLOBECOM) }, Kuala Lumpur, Malaysia, 2023, pp. 6493--6498, 

\bibitem{IRS-ISAC3}
Z. Wang, X. Mu, and Y. Liu, ``STARS Enabled Integrated Sensing and Communications,'' {\em IEEE Trans. Wireless Commun.}, vol. 22, no. 10, pp. 6750--6765, Oct. 2023.

\bibitem{IRS-ISAC4}
X. Song, D. Zhao, H. Hua, T. X. Han, X. Yang, and J. Xu, ``Joint transmit and reflective beamforming for IRS-assisted integrated sensing and communication,'' in {\em  Proc. IEEE Wireless Commun. Netw. Conf.}, Apr. 2022, pp. 189--194.

\bibitem{IRS-ISAC5}
T. Wei, L. Wu, K. V. Mishra, and M. R. B. Shankar, ``Multi-IRS-aided Doppler-tolerant wideband DFRC system,'' {\em IEEE Trans. Commun.}, vol. 71, no. 11, pp. 6561--6577, Nov. 2023.

\bibitem{TLS-ESPRIT}
X. Hu, C. Liu, M. Peng, and C. Zhong, ``IRS-based integrated location sensing and communication for mmWave SIMO systems,'' {\em IEEE Trans. Wireless Commun.}, vol. 22, no. 6, pp. 4132--4145, Jun. 2023.

\bibitem{mmWave}
Z. Yu, X. Hu, C. Liu, M. Peng, and C. Zhong, ``Location sensing and beamforming design for IRS-enabled multi-user ISAC systems,'' {\em IEEE Trans. Sig. Process.}, vol. 70, pp. 5178--5193, Nov. 2022

\bibitem{estimation3}
L. Xie, X. Yu, and S. Song, ``Intelligent reflecting surface-aided manoeuvring target sensing: True velocity estimation,'' in {\em Proc. Int. Symp. Wireless Commun. Syst.}, 2022, pp. 1--6.

\bibitem{estimation4}
D. Cong, S. Guo, S. Dang, and H. Zhang, ``Vehicular behavior-aware beamforming design for integrated sensing and communication systems,'' {\em IEEE Trans. Intell. Transp. Syst.}, vol. 24, no. 6, pp. 5923--5935, Jun. 2023.

\bibitem{estimation5}
Z. Liu, C. Yang, T. Zhou, and M. Peng, ``Performance model of terahertz joint radar-communication systems under random mobility,'' in {\em Proc. IEEE 95th Veh. Technol. Conf. (VTC-Spring)},
Jun. 2022, pp. 1--6.

\bibitem{basic}
P. Raviteja, K. T. Phan, Y. Hong, and E. Viterbo, ``Interference cancellation and iterative detection for orthogonal time frequency space modulation,'' {\em IEEE Trans. Wireless Commun.}, vol. 17, no. 10, pp. 6501--6515, Oct. 2018.

\bibitem{PAPR}
G. D. Surabhi, R. M. Augustine, and A. Chockalingam, ``Peak-to-average power ratio of OTFS modulation,'' {\em IEEE Commun. Lett.}, vol. 23, no. 6, pp. 999--1002, Jun. 2019.

\bibitem{OTFS-ISAC3}
L. Gaudio, M. Kobayashi, G. Caire, and G. Colavolpe, ``On the effectiveness of OTFS for joint radar parameter estimation and communication,'' {\em IEEE Trans. Wireless Commun.}, vol. 19, no. 9, pp. 5951--5965, Sept. 2020.

\bibitem{OTFS-ISAC7}
L. Gaudio, M. Kobayashi, G. Caire, and G. Colavolpe, ``Hybrid digitalanalog beamforming and MIMO radar with OTFS modulation,'' 2020, {\em arXiv:2009.08785}.

\bibitem{OTFS-ISAC2}
S. Li, W. Yuan, C. Liu, Z. Wei, J. Yuan, B. Bai, D. W. K. Ng, ``A novel ISAC transmission framework based on spatially-spread orthogonal time frequency space modulation,'' {\em IEEE J. Sel. Areas Commun.}, vol. 40, no. 6, pp. 1854--1872, Jun. 2022.

\bibitem{OTFS-ISAC1}
M. Keskin, C. Marcus, O. Eriksson, A. Alvarado, J. Widmer, and H. Wymeersch. ``Integrated sensing and communications with MIMO-OTFS,'' {\em arXiv preprint arXiv:2306.06361}, 2023.

\bibitem{OTFS-ISAC4}
W. Yuan, Z. Wei, S. Li, J. Yuan, and D. W. K. Ng, ``Integrated sensing and communication-assisted orthogonal time frequency space transmission for vehicular networks,'' {\em IEEE J Sel Top Signal Process}, vol. 15, no. 6, pp. 1515--1528, Nov. 2021.

%


\bibitem{IRS-OTFS5}
V. S. Bhat, G. Harshavardhan, and A. Chockalingam, ``Input-output relation and performance of RIS-aided OTFS with fractional Delay-Doppler,'' {\em IEEE Commun. Lett.}, vol. 27, no. 1, pp. 337--341, Jan. 2023.

\bibitem{IRS-OTFS1}
E. Basar, ``Reconfigurable intelligent surfaces for Doppler effect and
multipath fading mitigation,'' 2019, {\em arXiv:1912.04080}. [Online]. Available: http://arxiv.org/abs/1912.04080.

\bibitem{SBL_MUSIC}
J. Francis and V. P. Reddy, ``Delay-Doppler channel estimation in OTFS systems using DoA estimation techniques,'' in {\em Proc. IEEE VTC2022-Spring.}, Helsinki, Finland, Jun. 2022, pp. 1--5.

\bibitem{IRS-IASC-vehicular4}
K. Meng, Q. Wu, W. Chen and D. Li, ``Sensing-assisted communication in vehicular networks with intelligent surface,’’ {\em IEEE Trans. Veh. Technol.}, vol. 73, no. 1, pp. 876--893, Jan. 2024.

\bibitem{IRS-OTFS2}
M. Li, S. Zhang, Y. Ge, F. Gao, and P. Fan, ``Joint channel estimation and data detection for hybrid RIS aided millimeter wave OTFS systems,'' {\em IEEE Trans. Commun.}, vol. 70, no. 10, pp. 6832--6848, Oct. 2022.

\bibitem{IRS-OTFS3}
M. Li, S. Zhang, Y. Ge, F. Gao, and P. Fan, ``A novel transmission strategy for hybrid RIS aided millimeter wave OTFS systems,''in { \em 2022 14th International Conference on Wireless Communications and Signal Processing (WCSP)}, IEEE, 2022.

\bibitem{strongest}
X. Xu, L. Xiang, J. An, C. Dong, S. Sugiura, R. G. Maunder, L. Yang, and L. Hanzo, ``OTFS-aided RIS-assisted SAGIN systems outperform their OFDM counterparts in doubly selective high-doppler scenarios,''  {\em IEEE Internet Things J.}, vol. 10, no. 1, pp. 682--703, Jan. 2023.

\bibitem{IRS-OTFS7}
A. Thomas, K. Deka, S. Sharma, and N. Rajamohan, ``IRS-assisted OTFS system: Design and analysis,'' {\em IEEE Trans. Veh. Technol.}, vol. 72, no. 3, pp. 3345--3358, Mar. 2023.

\bibitem{IRS-IASC-vehicular3}
M. Li, S. Zhang, Y. Ge, Z. Li, F. Gao, and P. Fan, ``STAR-RIS aided integrated sensing and communication over high mobility scenario,’’ {\em IEEE Trans Commun.}, early access, Mar. 2024, doi: 10.1109/TCOMM.2024.3381725.

\bibitem{IRS-OTFS8}
A. S. Bora, K. T. Phan, and Y. Hong, ``IRS-assisted high mobility communications using OTFS modulation,'' {\em IEEE Wireless Commun. Lett.}, vol. 12, no. 2, pp. 376--380, Feb. 2023.

\bibitem{IRS-OTFS4}
Z. Li, W. Yuan, B. Li, J. Wu, C. You, and F. Meng, ``Reconfigurable-intelligent-surface-aided OTFS: Transmission scheme and channel estimation,''{\em IEEE Internet Things J.}, vol. 10, no. 22, pp. 19518--19532, 15 Nov.15, 2023.
SSS


\bibitem{embedded2}
Z. Wei, W. Yuan, S. Li, J. Yuan, and D. W. K. Ng, ``Off-grid channel estimation with sparse bayesian learning for OTFS systems,'' {\em IEEE Trans. Wireless Commun.}, vol. 21, no. 9, pp. 7407--7426, Sept. 2022.

\bibitem{hyper}
F. W. Olver, D. W. Lozier, R. F. Boisvert, and C. W. Clark, {\em NIST Handbook of Mathematical Functions}, New York: Cambridge University Press, 2010.

\bibitem{gamma}
A. Castano-Martinez and F. Lopez-Blazquez, Distribution of a sum of weighted noncentral chi-square variables, Commun. Stat., vol. 34, no. 3, pp. 515C524, Mar. 2005.

\bibitem{nakagami}
A. Papoulis, Probability, Random Variables, and Stochastic Process, 3rded. New York, NY, USA: McGraw-Hill, 1991.

\bibitem{pdf}
I. S. Gradshteyn and I. M. Ryzhik, {\em Table of Integrals, Series, and Products}, 5th ed. Orlando, FL, USA: Academic, 1994.

\bibitem{mimo1}
P. Wang, J. Fang, L. Dai, and H. Li, ``Joint transceiver and large intelligent surface design for massive MIMO mmWave systems,'' {\em IEEE Trans. Wireless Commun.}, vol. 20, no. 2, pp. 1052--1064, Feb. 2021.

\bibitem{subspace1}
X. Meng, F. Liu, S. Lu, S. P. Chepuri, and C. Masouros, ``RIS-assisted integrated sensing and communications: A subspace rotation approach,” in {\em Proc. IEEE Radar Conf. (RadarConf23)}, San Antonio, TX, USA, 2023, pp. 1--6.

\bibitem{subspace2}
F. Liu, Y. -F. Liu, A. Li, C. Masouros, and Y. C. Eldar, ``Cramér-Rao bound optimization for joint radar-communication beamforming,'' {\em IEEE Trans. Sig. Process.}, vol. 70, pp. 240--253, Dec. 2021.

\bibitem{CADMM}
K. Huang and N. D. Sidiropoulos, ``Consensus-ADMM for general quadratically constrained quadratic programming,'' {\em IEEE Trans. Sig. Process.}, vol. 64, no. 20, pp. 5297--5310, Oct. 2016.

\bibitem{LTE}
A. Monk, R. Hadani, M. Tsatsanis, and S. Rakib, ``OTFS - Orthogonal time frequency space: A novel modulation technique meeting 5G high mobility and massive MIMO challenges,'' 2016. Available online: https://arxiv.org/abs/1608.02993


\bibitem{asso1}
H. Hashida, Y. Kawamoto, N. Kato, M. Iwabuchi, and T. Murakami, ``Mobility-aware user association strategy for IRS-aided mm-Wave multibeam transmission towards 6G,'' {\em IEEE J. Sel. Areas Commun.}, vol. 40, no. 5, pp. 1667--1678, May. 2022.

\bibitem{asso2}
D. Jin, Y. Xiao, Y. Li, G. Shi, and D. Niyato, ``Optimizing intelligent reflecting surface-base station association for mobile networks,'' in {\em Proc IEEE Int. Conf. Commun.(ICC)}, Montreal, Canada, 2021, pp. 1--6.

\bibitem{asso3}
H. Zhang and H. Wei ``Analysis of intelligent reflecting surface-enhanced mobility through a line-of-sight state transition model'', 2403.07337v1, 2024. [Online]. Available: https://arxiv.org/html/2403.07337v1.

\bibitem{real-time-IRS}
Z. Huang, B. Zheng, and R. Zhang, ``Transforming fading channel from fast to slow: IRS-assisted high-mobility communication,'' in {\em Proc IEEE Int. Conf. Commun.(ICC)}, Montreal, QC, Canada, 2021, pp. 1--6.

\bibitem{errorbound}
G.~Nemes, ``Error bounds and exponential improvements for the asymptotic expansions of the gamma function and its reciprocal,'' \emph{Proceedings of the Royal Society of Edinburgh Section A: Mathematics}, vol.~145, no.~3, pp. 571--596, 2015.

\bibitem{MIMO-LB}
B. Ning, Z. Chen, W. Chen, and J. Fang, ``Beamforming optimization for intelligent reflecting surface assisted MIMO: A sum-path-gain maximization approach,'' {\em IEEE Wireless Commun. Lett.}, vol. 9, no. 7, pp. 1105--1109, 2020.

\bibitem{OFDM-sensing}
C. Sturm and W. Wiesbeck, ``Waveform design and signal processing aspects for fusion of wireless communications and radar sensing,'' Proceedings of the IEEE, vol. 99, no. 7, pp. 1236--1259, July 2011.

\bibitem{tao1}
Q. Tao, X. Hu, S. Zhang and C. Zhong, ``Integrated sensing and communication for symbiotic radio systems in mobile scenarios,'' {\em IEEE Trans. Wireless Commun.}, early access, doi: 10.1109/TWC.2024.3379653.

\bibitem{tao2}
Q. Tao, T. Xie, X. Hu, S. Zhang and D. Ding, ``Channel estimation and detection for intelligent reflecting surface-assisted orthogonal time frequency space systems,'' {\em IEEE Trans. Wireless Commun.}, early access, doi: 10.1109/TWC.2024.3349707.

\end{thebibliography}
\end{document}